\documentclass[a4paper,12pt]{article}
\usepackage{jheppub,esint,shuffle,psfrag}
\usepackage[utf8]{inputenc}
\usepackage{booktabs}

\usepackage{esint} 
\usepackage{breqn}
\usepackage{amsmath,bm}
\usepackage{hyperref}
\usepackage{mathrsfs} 
\usepackage{enumerate}

\usepackage{tikz-cd}

\usepackage{tikz}

\definecolor{myred}{RGB}{233, 33, 45}

\DeclareMathOperator{\tr}{\text{tr}}     
\newcommand{\vev}[1]{\big\langle #1 \big\rangle}
\newcommand\re[1]{(\ref{#1})}

\def\cG{  {\cal G}  }
\def\cO{  {\cal O}  }
\newcommand\lr[1]{{\left({#1}\right)}}
\newcommand \widebar [1] {\overline{#1}}

\def \qqquad {\qquad\quad}
\def \qqqquad {\qquad\qquad}
 
\newcommand{\cF}{{\cal F}}
\newcommand{\cN}{{\cal N}}

 \def\numberbysection{\@addtoreset{equation}{section}
                     \def\theequation{\thesection.\arabic{equation}}}

\title{
Correlation functions in four-dimensional superconformal long circular quivers}

\author[a]{Gregory P. Korchemsky} 
\author[a,b]{and Alessandro Testa}
\affiliation[a]{Institut de Physique Th\'eorique\footnote{Unit\'e Mixte de Recherche 3681 du CNRS}, Universit\'e Paris Saclay, CNRS,  91191 Gif-sur-Yvette, France}
\affiliation[b]{Dipartimento SMFI, Università di Parma and INFN Gruppo Collegato di Parma,
Viale G.P. Usberti 7/A, 43100 Parma, Italy}

\abstract{We study two- and three-point correlation functions of chiral primary half-BPS operators in four-dimensional $\mathcal{N}=2$ superconformal circular, cyclic symmetric quiver theories. Using supersymmetric localization, these functions can be expressed as matrix integrals which, in the planar limit, reduce to Fredholm determinants of certain semi-infinite matrices.  This powerful representation allows us to investigate the correlation functions across the parameter space of the quiver theory, including both weak and strong coupling regimes and various limits of the number of nodes and the operator scaling dimensions. At strong coupling, the standard semiclassical AdS/CFT expansion diverges in the long quiver limit. However, by incorporating both perturbative corrections (in negative powers of the 't Hooft coupling) and an infinite tower of nonperturbative, exponentially suppressed contributions, we derive a remarkably simple expression for the correlation functions in this limit. 
These functions exhibit exponential decay with increasing node separation and admit an interpretation within a five-dimensional effective theory. 
We determine the mass spectrum of excitations propagating along the emergent fifth dimension within this theory, finding it to be given by the zeros of Bessel functions.}

\begin{document}
\maketitle
 
\section{Introduction and summary}

In this paper, we study a class of four-dimensional $\mathcal{N}=2$ super Yang-Mills (SYM) theories known as  circular quivers. These theories, arising from orbifold projections of maximally supersymmetric $\mathcal{N}=4$ SYM, play an important role in the AdS/CFT correspondence and provide a rich framework for exploring non-perturbative phenomena through gauge-gravity dualities \cite{Aharony:1999ti}, integrability \cite{Beisert:2010jr}, and supersymmetric localization \cite{Pestun:2016zxk}.

\begin{figure}[h]
	\begin{center}
		\psfrag{L}[cc][cc]{$L$}\psfrag{1}[cc][cc]{$1$}\psfrag{2}[cc][cc]{$2$}\psfrag{3}[cc][cc]{$3$}
		\includegraphics[width=0.4\textwidth]{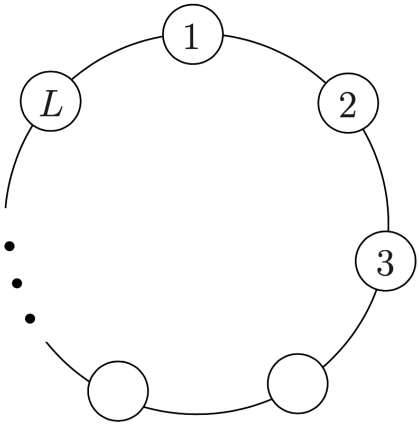}
		\caption{Diagrammatic representation of the circular quiver theory. Each node represents a $SU(N)$ vector multiplet, while lines connecting neighbouring nodes represent hypermultiplets in the bi-fundamental representation of $SU(N)\times  SU(N)$.}
		\label{fig:qiuv}
	\end{center}
\end{figure}

Their field content can be conveniently described by the quiver diagram shown in Fig.~\ref{fig:qiuv}, where each node represents a $SU(N)$ vector multiplet with coupling constant $g_I$ (with $I=1,\dots L$), and lines connecting neighbouring nodes denote massless hypermultiplets in the bi-fundamental representation of $SU(N)\times SU(N)$. These models  are conformally invariant at the quantum level for arbitrary values of the coupling constants $g_I$. Closely related to  $\mathcal{N}=4$ SYM, these theories provide a powerful setup  for developing and testing various non-perturbative approaches in $\mathcal{N}=2$ models 
\cite{Rey:2010ry,Mitev:2014yba,Fiol:2015mrp,Mitev:2015oty,Pini:2017ouj,Zarembo:2020tpf,Fiol:2020ojn,Billo:2021rdb,Ouyang:2020hwd,Galvagno:2020cgq,Beccaria:2021ksw,Billo:2022gmq,Billo:2022fnb}.
    
The relationship between quiver theories and $\mathcal{N}=4$ SYM becomes manifest when all the $L$ gauge couplings are equal, $g_1=\dots=g_L = g_{_{\rm YM}}$. The resulting $\mathcal{N}=2$ superconformal theory, denoted $\mathsf{Q}_L$, exhibits a symmetry under the cyclic shift of the nodes, $I \to I+1$, which significantly simplifies the properties of the model. For instance, in the planar limit, several observables -- including the free energy on a four-sphere  and circular half-BPS Wilson loop  -- match their $\mathcal{N}=4$ SYM counterparts for arbitrary 't Hooft coupling $\lambda = g_{\rm YM}^2 N$ \cite{Fiol:2020ojn,Zarembo:2020tpf,Beccaria:2021ksw,Beccaria:2023kbl}. However, despite these similarities, the two theories are not equivalent even in the planar limit. In fact, certain two- and three-point functions of half-BPS gauge-invariant operators differ, providing valuable probes of the relationship between these models \cite{Billo:2021rdb,Billo:2022fnb,Billo:2022lrv,Galvagno:2020cgq,Preti:2022inu}.

Our goal in this paper is to compute these correlation functions in planar $\mathsf{Q}_L$ theory for arbitrary values of the 't Hooft coupling $\lambda$ and the number of nodes $L$.
Another motivation for studying the circular quiver theories is that they are expected to exhibit novel properties in the limit of large number of nodes   \cite{Uhlemann:2019ypp,Uhlemann:2019lge,Coccia:2020cku,Coccia:2020wtk,Akhond:2022oaf,Nunez:2023loo,Beccaria:2023qnu,Sobko:2024ohd}.  
In this regime, the nodes of the quiver diagram in Fig.~\ref{fig:qiuv} become continuously distributed, suggesting an emergent higher-dimensional description.

The underlying mechanism, known as the deconstruction prescription, was originally proposed in \cite{Arkani-Hamed:2001kyx,Hill:2000mu} in the context of four-dimensional gauge theories defined over quiver diagrams analogous to those depicted in Fig.~\ref{fig:qiuv} in the limit $L\to\infty$. These theories are four-dimensional at high energies, while their low-energy effective action corresponds to a higher-dimensional theory with a discretized extra dimension. The scale separating the two regimes is determined by the vacuum expectation values  of the link fields. The same mechanism was subsequently applied to the $\mathcal{N}=2$ superconformal quiver theories \cite{Arkani-Hamed:2001wsh}. The emergence of the fifth dimension occurs on the Higgs branch, where all chiral multiplets acquire 
the same vacuum expectation value, thus breaking the conformal symmetry. The low-energy action takes the form of a five-dimensional Lorentz invariant supersymmetric gauge theory in which the effective lattice spacing in the fifth dimension is inversely proportional to the 't Hooft coupling and the expectation value of the chiral multiplets. Various quantitative checks of the deconstruction proposal were performed in \cite{Lambert:2012qy,Hayling:2017cva,Niarchos:2019onf}. 

In all the deconstruction examples above, the underlying four-dimensional gauge theories are strongly coupled, rendering explicit calculations intractable. To circumvent this problem, we investigate the deconstruction prescription in the $\mathcal{N}=2$ superconformal quiver theories in a different setting -- at the origin of the moduli space, for zero vacuum expectation values of the chiral multiplets. The main advantage of this choice is that localization enables us to compute various two- and three-point functions of half-BPS operators in the strong coupling regime for arbitrary number of nodes $L$. Examining the behaviour of these functions in the limit $L\to\infty$, we can identify an effective five-dimensional theory describing this regime. 

The emergent five-dimensional theory differs drastically from the one originating on the Higgs branch. The main reason is that the four-dimensional conformal symmetry remains unbroken in the limit $L\to\infty$. This symmetry imposes tight constraints on the form of correlation functions of local operators $O_{n_i}^{I_i}(x_i)$ defined at different nodes of the quiver. These correlation functions depend on both space-time separations of operators, $x_{ij}^2 = (x_i - x_j)^2$, and their separations on the quiver diagram, $Y_{ij}^2 = (I_i - I_j)^2$ (see Fig.~\ref{fig:qiuv}). 
Conformal symmetry dictates  that the two- and three-point correlators  factorize into a product of two independent functions, each depending separately on $x_{ij}^2$ and $Y_{ij}^2$. This factorization precludes Lorentz invariance in the effective five-dimensional theory. 

In the present work, we 
compute  two- and three-point functions of chiral primary operators $T_{\alpha,n}(x)$ in the planar ${\sf Q}_L$ theory and examine their behaviour in the limit $L\to\infty$. These operators have the following form
\begin{align}\label{T-def}
T_{\alpha,n}(x) =\dfrac{1}{\sqrt{L}}
\sum_{I=1}^{L} {e}^{-ip_\alpha I}  O_n^I(x)\,,
\end{align}
where $p_\alpha=2\pi\alpha/L$ (with $\alpha=0,\dots,L-1$)  and $O_n^I(x) = \tr \phi^n_I(x)$ is a single-trace operator built out of a scalar field $\phi_I(x)$ belonging to the vector multiplet at node $I$. The operators \re{T-def}  carry a $U(1)$ charge $n\ge 2$ and their scaling dimension $\Delta_{\alpha,n}=n$ is protected by supersymmetry. 
The relation \re{T-def} takes the form of a discrete Fourier transform on the quiver lattice with $p_\alpha$  having the meaning of a quasimomentum. 
In the long quiver limit, $p_\alpha$ becomes the momentum of excitations propagating along the fifth dimension. 

Being combined together, the conformal and cyclic symmetry of the ${\sf Q}_L$ theory  fix the form of two- and three-point correlation functions of the operators \re{T-def} as
\begin{subequations}
	\label{eq:2p intro}
	\begin{align}\label{G2}
		{}& \langle T_{\alpha_1,n_1}({x_1}) \widebar{T}_{\alpha_2,n_2}(x_2) \rangle = G_2
		\frac{\delta_{n_1,n_2} \delta(p_{\alpha_1}+p_{\alpha_2})}{|x_1-x_2|^{2n_1}} 
		 \,, \\[0.4em]\label{G3}
		{}&	\langle T_{\alpha_1,n_1}({x_1}) T_{\alpha_2,n_2}(x_2) \widebar{T}_{\alpha_3,n_3}(x_3)\rangle  =  G_3 {\delta_{n_1+n_2,n_3} \delta(p_{\alpha_1}+p_{\alpha_2}+p_{\alpha_3})\over |x_1-x_3|^{2n_1}|x_2-x_3|^{2n_2}} \,,
	\end{align}
\end{subequations}
where the operator $\widebar{T}_{\alpha,n}(x)=T^\dagger_{L-\alpha,n}(x)$
is obtained from \re{T-def} by replacing $O_n^I$ with the complex conjugated operator. 
The product of delta-functions in \re{eq:2p intro} arises from the conservation of the $U(1)$ charge and the quasimomentum.~\footnote{For the quasimomenta, the delta-functions in \re{eq:2p intro} fix their values modulo $2\pi$.}
The normalization factors $G_2$ and $G_3$ encode the dependence of the correlation functions on the quantum numbers of the operators, $n_i$ and $p_{\alpha_i}$, as well as on the parameters of the ${\sf Q}_L$ theory -- the 't Hooft coupling $\lambda$, the number of nodes $L$ and the gauge group rank $N$.

For the operators \re{T-def} with zero quasimomentum, $p_\alpha=0$, their two- and three-point functions \re{eq:2p intro} are protected by supersymmetry \cite{Billo:2021rdb,Billo:2022fnb,Billo:2022lrv}. The corresponding normalization factors $G_2$ and $G_3$ are independent of the coupling constant and coincide with their counterparts in $\mathcal N=4$ SYM. For the operators \re{T-def} carrying nonzero quasimomentum  $p_\alpha\neq 0$, the correlation functions \re{eq:2p intro} can be computed for arbitrary coupling using the  supersymmetric localization, which maps the normalization factors $G_2$ and $G_3$ to 
correlators in an interacting matrix model (see (\ref{eq:partition function on S4}) and (\ref{OO-R}) below). 
Evaluation of these correlators in the planar limit, for $N\to\infty$ with $\lambda=g^2_{_{\rm YM}}N$ held fixed, shows that both factors can be expressed in terms of the same function, schematically, \cite{Billo:2022lrv}
\begin{align}\label{G-to-R}
G_2\sim R_{\alpha,n} \,,\qqqquad G_3\sim  \prod_{i=1}^{3} \sqrt{ (n_i+ \lambda\partial_\lambda) R_{\alpha_i,n_i}} \,.
\end{align}
Most importantly, the function $R_{\alpha,n}$ admits a closed form representation in terms of Fredholm determinants of certain (infinite-dimensional) integrable Bessel operators (see  (\ref{eq:twisted two-point})). 
The relation \re{G-to-R} holds for arbitrary 't Hooft coupling  $\lambda$ and the number of nodes $L$. This provides a framework for studying the correlation functions \re{eq:2p intro} within  the parameter space of the quiver theory. 

Following the approach of \cite{Beccaria:2023qnu}, we interpret the quiver diagram in Fig.~\ref{fig:qiuv} as a one-dimensional lattice model whose partition function coincides with that of the interacting matrix model obtained via localization. This correspondence allows us to understand properties of the quiver theory through those of the lattice model. In particular, 
the leading nonplanar correction to the free energy of the ${\sf Q}_L$ theory defined on the unit four-sphere equals the free energy of this model. Its dependence on the number of nodes $L$ was studied in \cite{Beccaria:2023qnu}, where it was shown that the free energy has different asymptotic behaviour  depending on how $L$ compares with the coupling constant. We demonstrate below that the same phenomenon happens for the correlation functions \re{eq:2p intro}.

In the lattice model description, the two-point correlation function \re{G2} describes a propagation of $n$-particle excitation carrying the quasimomentum $p_\alpha=2\pi\alpha/L$. In the similar manner, the three-point function \re{G3} describes the transition amplitude $1+2\to 3$, where $i$ denotes an excitation containing $n_i$ particles  with the total quasimomenta $p_{\alpha_i}$.
 
In the weak coupling regime, the two-point function $G_2\sim R_{\alpha,n}$ is given by a double series in 't Hooft coupling $\lambda$ and the parameter $s_\alpha=\sin^2\left( {p_\alpha}/{2}\right)$ which depends on the quasimomentum. Performing a resummation of the leading terms in this series 
we get
\begin{equation}\label{R-resum0}
 		R_{\alpha,n} ={1\over 1+ r_n\, \lambda^n \sin^2\left( {p_\alpha}/{2}\right)} + \dots \, ,
\end{equation} 
where dots denote subleading corrections and the coefficient $r_n$ is proportional to odd Riemann zeta-value. A distinguished feature of \re{R-resum0} is that it develops poles for complex $p_\alpha$. Note that these poles are not present at any fixed order of the weak coupling expansion of $R_{\alpha,n}$. The closest to origin pole is located at $p_\alpha=\pm i \mu_n$, where $\mu_n\sim n\log(1/\lambda)$ grows logarithmically at weak coupling. 

The appearance of poles in the two-point function \re{R-resum0} implies that the excitations propagating across the quiver diagram acquire nonzero masses. This suggests that the correlations between different sites in the lattice quiver model should be suppressed as a function of their separation. Indeed, performing the inverse discrete Fourier transformation of \re{R-resum0} we find that the correlation functions of the local operators $O_n^I(x)$ defined in \re{T-def}  decays exponentially with the node separation
\begin{align}\label{OO-exp}
\vev{O_n^I(x) O_n^J(0)} \sim {e^{-\mu_n |I-J|}\over x^{2n}} \,.
\end{align}
This relation holds at weak coupling in the limit $L\to\infty$ with $|I-J|$ kept fixed. We deduce from \re{OO-exp} that the one-dimensional quiver lattice model exhibits an extensive behaviour and the interaction between the sites is of short range.  This agrees with the results of  \cite{Beccaria:2023qnu}, where it was shown that the free energy of the model scales linearly with its length $L$ at weak coupling.   

As mentioned earlier, the behaviour of the correlation functions \re{eq:2p intro} at strong coupling crucially depends on the ratio $\sqrt{\lambda}/L$. At large $L$ and $\sqrt\lambda\gg L$, the correlation functions \re{eq:2p intro} are expected to admit a semiclassical expansion in powers of $1/\sqrt\lambda$ within the AdS/CFT correspondence. Indeed, exploiting the relation between the function $R_{\alpha,n}$ entering \re{G-to-R} and the Fredholm determinants of the Bessel kernel, we obtain
\begin{align}\label{R-str} 
 R_{\alpha,n}={(n-1) n\over \xi^2}  \left[1 +f_0\lr{1/ \xi} + e^{-2\xi} f_1\lr{1/ \xi}    +O(e^{-4\xi}) \right],
\end{align}
where $\xi=p_\alpha \sqrt\lambda /(4\pi)$. The relation \re{R-str} takes the form of a transseries with the coefficient functions $f_0$ and $f_1$ given by series in $1/\xi$. These functions describe, respectively, the perturbative and the leading nonperturbative corrections to \re{R-str} at strong coupling.

In the limit $\sqrt{\lambda}\gg L\gg1$ (or equivalently $\xi\gg 1$), the non-perturbative corrections to \re{R-str} become exponentially small and, as expected, the function $R_{\alpha,n}$ is given by series in $1/\sqrt{\lambda}$ with the coefficients depending on the quiver length (see \re{R-simp} and (\ref{small-xi})). In this case, all sites of the one-dimensional lattice model interact equally strongly and the very notion of site ordering along the lattice becomes meaningless. Consequently, the lattice effectively collapses to a point and the quiver theory remains a four-dimensional, strongly coupled theory without an emergent dimension.

In the opposite limit, for $L \gg \sqrt{\lambda} \gg 1$ (or equivalently $\xi \ll 1$), the semiclassical expansion \re{R-str} breaks down  \cite{Beccaria:2023qnu}. Determining $R_{\alpha,n}$ in this regime requires not only resumming perturbative corrections to $f_0(1/\xi)$ but also account for an infinite tower of non-perturbative corrections to \re{R-str}. We achieved this goal by using special properties of the Fredholm determinants of the Bessel operators \cite{Beccaria:2022ypy,Beccaria:2023qnu,Bajnok:2024epf,Bajnok:2024ymr}. We show that for arbitrary $\xi$ the function \re{R-str}  can be elegantly expressed in terms of modified Bessel functions
\begin{align}\label{R-resum}  
R_{\alpha,n}={(n-1) n\over \xi ^{2} } \,\frac{ {\mathrm I}_n(2\xi)}{{\mathrm I}_{n-2}(2\xi)}\,.
\end{align} 
This relation holds for $\lambda$ and $L$ going to infinity with $\xi=\alpha\sqrt\lambda /(2L)$ kept fixed. 
Expanding the right-hand side of \re{R-resum} at large $\xi$ we recover the relation \re{R-str}. 

According to \re{R-resum}, $R_{\alpha,n}$ is a meromorphic function of the quasimomentum $p_\alpha$. It takes positive values for real $p_\alpha$ and has an infinite number of poles along the imaginary axis at $p_\alpha=\pm 2\pi i m_k/\sqrt\lambda$ (with $k\ge 1$) coming from zeros of the Bessel function in the denominator of \re{R-resum}. These poles satisfy $J_{n-2}(m_k)=0$ and their position depends on the scaling dimension of the operator $n$. The function \re{R-resum} can be expanded into the sum over its poles
\begin{align}\label{R-masses} 
R_{\alpha,n}={64\pi^2 (n-1)^2 n\over\lambda} \sum_{k=1}^{\infty}\frac{f_k}{ p_\alpha^2+(2\pi m_{k}/\sqrt\lambda)^2} \,,
\end{align} 
where $f_k=1/m_k^2$. The right-hand side contains the sum of one-dimensional Euclidean scalar propagators of particles with mass $2\pi m_{k}/\sqrt\lambda$.

Being combined together, the relations \re{G-to-R} and \re{R-masses} determine the two- and three-point correlation functions \re{eq:2p intro} in the long quiver limit $L\gg \sqrt\lambda \gg 1$. The resulting expressions for the functions $G_2$ and $G_3$ admit a straightforward interpretation within a one-dimensional effective theory.
In fact, the relation \re{R-masses} takes the form of a K\"all\'en-Lehmann representation of the two-point correlation function in this effective description. 
The excitations described by \re{R-masses} propagate across the quiver diagram and interact with each other through a cubic vertex defined by the three-point function $G_3$.
 
Importantly, all masses in \re{R-masses} are non-zero. This property, analogous to the weak coupling behavior \re{OO-exp}, ensures the exponential decay of correlation functions at large distances along the quiver. The dominant contribution to (\ref{R-masses}) comes from the excitations with the minimal mass. Consequently, the characteristic mass scale in \re{OO-exp} at strong coupling is given by $\mu_n = 2\pi m_1 / \sqrt{\lambda}$. The inverse mass, $1/\mu_n$, defines the correlation length. Consistent with our expectations, this correlation length grows as $\sqrt{\lambda}$ in the strong coupling regime.
  
This paper is organized as follows. In Section \ref{sec2}, we summarize the properties of the correlation functions (\ref{eq:2p intro}) and discuss their connection with the Fredholm determinants of certain integrable Bessel operators. In Section \ref{sec:different regimes}, we discuss the different regimes in the parameter space of the quiver theory and define the long quiver limit. Section \ref{sect4} is devoted to the study of the correlation functions in the weak coupling regime. In Section \ref{sec:Sec5} we derive the strong coupling expansion of the correlation functions \re{eq:2p intro} and discuss their behaviour in different regimes. The long-quiver limit of the correlation function at strong coupling is studied In Section~\ref{sect6}.
Its interpretation in terms of effective five-dimensional theory is discussed in Section \ref{sec7}. The concluding remarks are presented in Section~\ref{sect8}.

\section{Correlation functions in circular quiver theories}\label{sec2}

The primary goal of this paper is to compute the two- and three-point functions of the conformal primary operators \re{T-def} in four-dimensional $\mathcal{N}=2$ superconformal circular quiver theory represented in Figure~\ref{fig:qiuv}, in the planar limit and for arbitrary number of nodes $L$. 

As outlined in the Introduction, we assume that the coupling constants at all nodes are equal to $g_{_{\rm YM}}$, thereby ensuring that the theory enjoys the cyclic symmetry.
Consequently, the conformal primary operators $T_{\alpha,n}(x)$ and $\widebar T_{\alpha,n}(x)$ are given by linear combinations of  single-trace (anti) chiral operators defined locally at each node $I$ of the quiver diagram \cite{Billo:2021rdb,Billo:2022fnb}
\begin{align}\label{eq:chiral operators} 
O_n^I(x) = \tr \phi^n_I(x)\,, \qqqquad  
		\widebar{O}_n^I(x)&=\tr \phi_I^{\dagger\,n}(x) \,,
\end{align}
where $\phi_I(x)$ and its complex conjugate $\phi^\dagger_I(x)$ are scalar fields of the $I$-th vector multiplet taking values in the $su(N)$ Lie algebra. 

Due to the interaction between different nodes of the quiver, the operators \re{eq:chiral operators} mix with the operators at other nodes carrying the same $U(1)$ charge and scaling dimension. 
In virtue of the cyclic symmetry of the quiver, the conformal primary operators $T_{\alpha,n}(x)$  belong to a representation of the $\mathbb Z_L$ group and acquire a phase 
under the cyclic shift, $\phi_I\to \phi_{I+1}$,  
\begin{equation}\label{shift}
 T_{\alpha,n}(x)\to  {e}^{\frac{2\pi i\alpha }{L} } T_{\alpha,n}(x) \,, 
\end{equation} 
where $\alpha=0,\dots,L-1$. This condition fixes the relative coefficients in the expansion \re{T-def}. 

Depending on the corresponding value of the quasimomentum $p_\alpha=2\pi\alpha/L$, it is convenient to distinguish {\it untwisted} ($p_\alpha=0$) and {\it twisted} ($p_\alpha \neq 0$) conformal primary operators. A distinguishing feature of the untwisted operators, $T_{0,n}(x)$ and $\widebar T_{0,n}(x)$, is that they exhibit the same properties of half-BPS scalar operators in  $\mathcal N=4$ SYM. Specifically, their two- and three-point correlation functions \re{eq:2p intro} are protected, meaning that the normalization factors $G_2$ and $G_3$ 
are independent of the coupling constant for $\alpha_i=0$. In contrast, for $\alpha_i \neq 0$, they are nontrivial functions of the 't Hooft coupling. 

Inverting the relation \re{T-def}, we can apply \re{eq:2p intro} to compute the correlation functions of the operators \re{eq:chiral operators} 
\begin{subequations}
	\label{OO-corr}
\begin{align}\label{OO}
	{}&\vev{O_n^I(x)\widebar O_m^J(0)} =D_{IJ} { \delta_{nm}\over x^{2n}}\,,\\[0.4em]
	\label{OOObar ft}
	{}&	\langle O^I_{n_1}({x_1}) O^J_{n_2}(x_2) \widebar{O}^K_{n_3}(x_3)\rangle  =  D_{IJK}{\delta_{n_1+n_2,n_3} \over |x_1-x_3|^{2n_1}|x_2-x_3|^{2n_2}} \,.
\end{align} 
\end{subequations}
The normalization factors $D_{IJ}$ and $D_{IJK}$ depend on the distances between the nodes and are related to the factors $G_2$ and $G_3$, introduced in \re{eq:2p intro},  by a discrete Fourier transformation \re{T-def}. 

The relations \re{eq:2p intro} and \re{OO-corr} are equivalent. As we show below, they play a complementary role in understanding the properties of the quiver theory in the limit of large number of nodes. 
The advantage of \re{eq:2p intro} is that, for arbitrary values of the parameters of the quiver theory, the normalization factors $G_2$ and $G_3$ can be computed using supersymmetric localization in terms of matrix integrals \cite{Billo:2022fnb,Billo:2022lrv}.

\paragraph{Two-point and three-point functions from localization}

Supersymmetric localization \cite{Pestun:2007rz} can be applied to compute the correlation functions \re{eq:2p intro} in terms of  the following interacting matrix model 
\begin{equation}
	\label{eq:partition function on S4}
	\mathcal{Z} = \int\prod_{I=1}^{L}\mathcal{D}a_I \exp\left(-\sum_{I=1}^{L}\left[\tr a_I^2-S_{\rm int}(a_I,a_{I+1})\right]\right),
\end{equation}
where the integration goes over ${su}(N)$ matrices $a_I$, subjected to the periodicity condition $a_I=a_{I+L}$. At large $N$, the interaction potential in \re{eq:partition function on S4} is given by an infinite sum of double traces \cite{Pini:2017ouj,Fiol:2020ojn,Billo:2021rdb,Billo:2022fnb,Beccaria:2023qnu}
\begin{equation}
	\label{eq:interaction potential}
	S_{\rm int}(a_I,a_{I+1})= \sum_{i,j=2}^{\infty}C_{ij}
	\left(\tr  {a}_I^i-\tr  {a}_{I+1}^{i}\right)\left(\tr  {a}_I^j-\tr  {a}_{I+1}^{j}\right)\,, 
\end{equation}
where the coefficients $C_{ij}$ are given by the product of powers of the coupling constant and Riemann zeta values. 
Their explicit expressions are irrelevant for our purposes.
The relation \re{eq:partition function on S4} yields the partition function of the quiver theory $\mathsf{Q}_L$  on a four-dimensional sphere of unit radius. A detailed analysis of the free energy $F=-\log \mathcal{Z}$ and its dependence on 't Hooft coupling and the number of nodes $L$ is discussed in \cite{Beccaria:2023qnu}. 

The correlation functions (\ref{eq:2p intro}) can be computed as expectation values of single traces in the lattice matrix model \re{eq:partition function on S4}.
Specifically, in the matrix model representation the conformal primary operators (\ref{T-def}) are mapped to linear combinations of single traces \cite{Gerchkovitz:2016gxx,Rodriguez-Gomez:2016ijh}
\begin{align}
\mathcal O_{\alpha,n} = \sum_I \sum_{2\le k\le n} \mathrm{e}^{-\frac{2\pi i I}{L}\alpha} c_k \tr a_I^k\,.
\end{align}
The
expansion coefficients $c_k$ are determined by the condition that the correlator $\vev{\mathcal O_{\alpha,n} \widebar{\mathcal O}_{\beta,m} }$ has to be 
diagonal with respect to both indices
\begin{align}\label{OO-R}
	\vev{\mathcal O_{\alpha,n} \widebar{\mathcal O}_{\beta,m} } = G_{\alpha,n} \delta_{nm}\delta_{\alpha\beta}\,.
\end{align}
Computing these correlators in the matrix model \re{eq:partition function on S4}, we obtain the following expression for the normalization factor of the two-point function  \re{eq:2p intro} in large $N$ limit 
\cite{Billo:2021rdb,Billo:2022fnb} 
\begin{equation}
	\label{eq:coefficients}
G_2\equiv  G_{\alpha,n}= {R}_{\alpha,n}(g)\,\cG_n\,,  
\end{equation} 
where  $0\le \alpha\le L-1$. In this relation, the $SU(N)$ color factor $\cG_n$ and the coupling constant $g$ are defined as
\begin{align}
	\cG_n= n\left(\frac{N}{2}\right)^n,\qqqquad g=\frac{\sqrt{\lambda}}{4\pi} \,.
\end{align}
In what follows, we find it convenient to employ the  coupling constant $g$ instead of  $\lambda$ in order to simplify expressions.

The coupling dependent function ${R}_{\alpha,n}(g)$ has different properties for $\alpha=0$ and $\alpha\neq 0$.  As noted earlier, the correlation function of untwisted operators, for $\alpha=0$, is independent of the coupling constant and coincides with its counterpart in $\mathcal{N}=4$ SYM. As a consequence, $R_{0,n}=1$ and $G_2=\cG_n$ matches the two-point correlation function of half-BPS   operators in planar $\mathcal{N}=4$ SYM \cite{Billo:2021rdb,Billo:2022fnb,Billo:2022lrv}. 
Conversely, for $\alpha\neq 0$, the function ${R}_{\alpha,n}(g)$ in \re{eq:coefficients} admits a representation in terms of Fredholm determinants of certain semi-infinite matrices (see \re{eq:twisted two-point} below)  \cite{Billo:2022fnb,Beccaria:2022ypy}.
 
For the three-point functions \re{G3}, supersymmetric localization provides the expression of the normalization factor in terms of a three-point correlator in the matrix model \re{eq:partition function on S4}
\begin{align}\label{OOObar}
G_3=	\langle {\mathcal O}_{\alpha_1,n_1} {\mathcal O}_{\alpha_2,n_2} \widebar{{\mathcal O}}_{\alpha_3,n_3} \rangle\,,
\end{align}
where  $n_3=n_1+n_2$ and $\alpha_1+\alpha_2+\alpha_3=0$ (mod $L$). It worth mentioning that the correlation functions \re{eq:2p intro} depend on the normalization of the operators. 
To remove this dependence, we introduce the ratio
\begin{equation}\label{C123}
 C_3= \frac{G_3}{\sqrt{  
 G_{{\alpha_1,n_1}} G_{{\alpha_2,n_2}}G_{{\alpha_3,n_3}}
}} \,.
\end{equation}
It depends on the three sets of quantum numbers $(\alpha_i,n_i)$ and defines the OPE coefficients of the operators \re{T-def}.

Computing the correlator \re{OOObar} in the large $N$ limit yields the following result \cite{Billo:2022lrv} 
\begin{align}\label{eq:strcutre constant def}
C_3 = \dfrac{\sqrt{n_1 n_2 n_3}}{\sqrt{L}N} \mathcal{V}_{\alpha_1,n_1}(g)\mathcal{V}_{\alpha_2,n_2}(g)\mathcal{V}_{\alpha_3,n_3}(g)\,.
\end{align}
This relation is valid in the leading large $N$ limit. 
The expression on the right-hand side factorizes into the product of three functions, one for each operator. Remarkably, the function ${\mathcal{V}}_{\alpha,n}$ can be expressed in terms of the two-point function \re{eq:coefficients}
\begin{align}\label{V}
	{\mathcal{V}}_{\alpha,n} = \sqrt{1+\frac{1}{2n}g\partial_g\log  {R}_{\alpha,n} } \,.
\end{align}
For $\alpha=0$ we have ${R}_{0,n}=1$ and the corresponding function ${\mathcal{V}}_{0,n}$ ceases to depend on 't Hooft coupling, as expected for the untwisted operators.
The relations \re{eq:strcutre constant def} and \re{V} were first derived  in \cite{Billo:2022lrv} for  $L=2$ quiver theories and then generalized to arbitrary $L$.

\paragraph{Relation to Fredholm determinants}  According to \re{OO-R} and \re{eq:coefficients}, the function $R_{\alpha,n}$ it is given by the two-point correlator in the matrix model \re{eq:partition function on S4}.  Expanding the integral in \re{eq:partition function on S4} in powers of $S_{\rm int}$ and performing a Gaussian integration, we find that all terms in the interaction potential \re{eq:interaction potential} contribute to $R_{\alpha,n}$. 
In the large $N$ limit,   $R_{\alpha,n}$ admits two equivalent representations  \cite{Billo:2022fnb,Beccaria:2022ypy}
\begin{align}
	\label{eq:twisted two-point} 
	{R}_{\alpha,n}(g) =\lr{1\over 1-s_\alpha K_{n-1}(g) }_{11} = \dfrac{\det(1-s_\alpha K_{n+1}(g) )}{\det(1-s_\alpha K_{n-1}(g) )} \,,
\end{align} 
involving a semi-infinite matrix $K_{n-1}(g)$ defined below.
The first relation contains the top-left entry of the inverse matrix $(1-s_\alpha K_{n-1}(g) )^{-1}$. In the second relation, the matrix $K_{n+1}$ is obtained from $K_{n-1}$ by removing its first row and column.
The equivalence of the two representations \re{eq:twisted two-point} follows from Cramer's rule.

In the relation \re{eq:twisted two-point},  the semi-infinite matrices $K_{n+1}$ and $K_{n-1}$ belong to a one-parameter family of  Bessel matrices $\left(K_\ell\right)_{ij} $ defined as 
 (for $i,j\ge 1$)
\begin{align}
	\label{eq:matrix}
	\left(K_\ell\right)_{ij} &=(-1)^{i+j}\sqrt{2i+\ell-1}\sqrt{2j+\ell-1}\int_0^\infty\dfrac{d t}{t}J_{2i+\ell-1}(\sqrt{t})J_{2j+\ell-1}(\sqrt{t})\chi\left(\frac{\sqrt{t}}{2g}\right)\,,
\end{align}
where  $J_m(x)$ is a Bessel function of the first kind (hence the name of the matrix) and $\chi(x)$ is a rapidly decreasing function at infinity
\begin{align}\label{eq:symbol}
	\chi(x) =-\dfrac{1}{\sinh^2\left(\frac{x}{2}\right)}\,.
\end{align}
The properties of this function, conventionally called \textit{symbol} of the matrix, play an important  role in our analysis.

The dependence of ${R}_{\alpha,n}(g)$ on $\alpha$ and $L$ enters \re{eq:twisted two-point} through the function
\begin{align}
	\label{eq:quantities}
	s_\alpha&=\sin^2\left(\frac{\pi\alpha}{L}\right) \,.
\end{align}
The vanishing of $s_\alpha$ for $\alpha=0$ implies ${R}_{0,n}(g)=1$, in agreement with the expected independence of the correlation functions \re{eq:coefficients} of untwisted operators  from the 't Hooft coupling. Due to the invariance of $s_\alpha$ under $\alpha\to L-\alpha$ and $\alpha\to-\alpha$, the function ${R}_{\alpha,n}(g)$ has the same property, 
\begin{align}\label{R-inv}
	{R}_{\alpha,n}(g)= {R}_{L-\alpha,n}(g)= {R}_{-\alpha,n}(g)\,.
\end{align}

The matrix \re{eq:matrix} is closely related to the semi-infinite matrix $C_{ij}$ in \re{eq:interaction potential}. Specifically, the expansion of the ratio  \re{eq:twisted two-point} in powers of $K_{n+1}$ and $K_{n-1}$ is in one-to-one correspondence with the expansion of the correlator 
\re{OO-R} in the matrix model \re{eq:partition function on S4} in powers of $S_{\rm int}$. Importantly, the relation \re{eq:twisted two-point} holds for arbitrary values of parameters
$n$, $L$ and $g$. We will use it to examine the properties of the correlation functions for different values of these parameters.

At finite coupling, the determinants in \re{eq:twisted two-point} can be analyzed using the properties of the 
matrix \re{eq:matrix}. As was shown in \cite{Beccaria:2023qnu,Beccaria:2022ypy}, this matrix is closely related to the so-called integrable Bessel kernel. Its Fredholm determinant
\begin{align}\label{F}
	e^{\mathcal F_\ell} =\det(1-s_\alpha K_{\ell} )
\end{align}
can be identified as a generalized Tracy-Widom distribution. Exploiting this relationship, we  can show that $\mathcal F_\ell$ satisfies a Toda-like equation \cite{DS,GK}
\begin{align}\label{Toda}
	g\partial_g\lr{\mathcal F_{\ell+1}-\mathcal F_{\ell-1}}  = 2\ell \lr{\mathrm{e}^{2\mathcal{F}_{\ell} -\cF_{\ell-1} -\cF_{\ell+1}}-1}\,.
\end{align}
This relation holds for arbitrary values of the parameters.

Taking into account \re{F} and \re{Toda}, we can simplify  \re{eq:twisted two-point} and \re{V} as follows
\begin{align}\label{eq:toda equation}\notag
	{}& R_{\alpha,n} = \exp\lr{\mathcal F_{n+1}-\mathcal F_{n-1}}\,,
	\\[1.2mm]
	{}&  {\mathcal{V}}_{\alpha,n} =  \exp\lr{\mathcal{F}_{n} -\frac12\cF_{n-1} -\frac12\cF_{n+1} }\,,
\end{align} 
where the dependence of $\mathcal F_n$ on $\alpha$ is tacitly assumed.
Being combined with \re{eq:strcutre constant def}, these relations allow us to express the two-point functions \re{eq:coefficients}  and three-point functions \re{OOObar} in the ${\sf Q}_L$ theory as 
\begin{subequations}\label{C123-GF}
 \begin{align}
 {}& G_2=G^{(0)}_2\exp\lr{\mathcal F_{n+1}-\mathcal F_{n-1}}\,,
 \\[2mm]
 \label{G123-F}
{}& G_3=G^{(0)}_3\prod_{i=1}^3 \exp\lr{\mathcal{F}_{n_i} -\cF_{n_i-1} }\,,
 \\\label{C123-F}
{}& C_3= \dfrac{\sqrt{n_1 n_2 n_3}}{\sqrt{L}N}\prod_{i=1}^3 \exp\lr{\mathcal{F}_{n_i} -\frac12\cF_{n_i-1} -\frac12\cF_{n_i+1} }\,.
 \end{align}
 \end{subequations}
Here $G^{(0)}_n$ are the correlation functions for zero value of the coupling constant, their explicit form is not relevant to our analysis. The expressions on the right-hand side of \re{G123-F} and \re{C123-F} are invariant under the interchange of any pair of $(\alpha_i, n_i)$. Although they are well-defined for arbitrary $\alpha_i$ and $n_i$, we are interested in the specific case where $n_3 = n_1 + n_2$ and $\alpha_1 + \alpha_2 + \alpha_3 \equiv 0 \pmod{L}$.
 
\section{Properties of the circular quivers}
\label{sec:different regimes}

The correlation functions \re{eq:2p intro} depend non-trivially on the coupling constant
(\ref{eq:coefficients}), the number of nodes $L$ and the quantum numbers of the operators $(\alpha_i,n_i)$. In the representation  \re{C123-GF} this dependence is encoded in the properties of the Fredholm determinants \re{F}. 

\subsection{Different regimes}

We demonstrate below that the functions \re{C123-GF} exhibit different behaviour depending on the hierarchy of the aforementioned parameters. We can gain insight into these different regimes by examining the two-point function  
$G_2\sim R_{\alpha,n}(g)$ at weak and strong coupling. 

At weak coupling, we can apply \re{eq:twisted two-point} and expand the determinants in powers of the matrices 
$K_{n-1}$ and $K_{n+1}$. In this way, we can show that (see eq.(\ref{eq:perturbative expansio twisted}) below)
\begin{equation}
	\label{eq:expansion at weak}
	\begin{split}
		{R}_{\alpha,n}(g) =	 1 -4g^{2n} s_\alpha  \binom{2n}{n}\left[\zeta(2n-1)-g^{2} \frac{4 n (2 n+1)}{n+1} \zeta(2n+1)+ \cO(g^{4})\right] \,,		
	\end{split}
\end{equation} 
where the dependence on the quiver length $L$ enters via $s_\alpha$ defined in (\ref{eq:quantities}). 

Analogously, at strong coupling, we apply the large $g$ expansion of the Fredholm determinant \re{F} to find (see \re{RV} below)
\begin{equation}
	\label{eq:expansion at strong}
	R_{\alpha,n}(g)= \dfrac{n(n-1)}{4g^2 s_\alpha} \left[ 1-\frac{(n-1)}{2 g}I_1 +\frac{(n-2) (n-1)}{8
		g^2}I_1^2 +\dots\right] \,,
\end{equation} 
where the dots stand for subleading corrections. The expansion coefficients depend on the function $I_1=I_1(\alpha/L)$ defined in \re{In} below. 

Examining the relations \re{eq:expansion at weak} and \re{eq:expansion at strong}, we observe that the expansion coefficients within $R_{\alpha,n}(g)$ grow at large $n$ and/or small $\alpha/L$. This allows us to distinguish three different regimes corresponding to possible values of $n$ and $a=\alpha/L$
\begin{align}\label{regimes}
	\text{(i)\quad fixed $n$ and $a$}\,; \qqquad 
	\text{(ii)\quad $a\to 0$ with $n$ fixed}\,; \qqquad 
	\text{(iii)\quad $n\gg 1$ with $a$ fixed}\,.
\end{align}
In the first regime, the function $R_{\alpha,n}(g)$ is described by (\ref{eq:expansion at weak}) at weak coupling and by (\ref{eq:expansion at strong}) at strong coupling. By merging these expansions, we can determine $R_{\alpha,n}(g)$ for arbitrary coupling.

Regime (ii) corresponds to the long quiver limit, $L \to \infty$. As $\alpha/L \to 0$, the weak-coupling expansion coefficients in (\ref{eq:expansion at weak}) vanish, while the strong-coupling expansion coefficients in (\ref{eq:expansion at strong}) develop poles in $\alpha/L$ due to $I_1 \sim \alpha/L$. Consequently, the strong-coupling expansion (\ref{eq:expansion at strong}) diverges. Determining the correlation function in this regimes requires resumming perturbative corrections, which are power-suppressed in $1/g$, and non-perturbative contributions, which are exponentially small at large $g$.

Finally, regime (iii) corresponds to infinitely heavy operators, $n\to\infty$. In this limit, as in the previous case, the perturbative corrections to (\ref{eq:expansion at weak}) vanish as $O(g^{2n})$ at weak coupling, while the expansion coefficients in (\ref{eq:expansion at strong}) grow as powers of $n$ at strong coupling. To determine the correlation function in this limit, we have to perform a resummation of the strong coupling expansion  (\ref{eq:expansion at strong}) to all orders in $1/g$.

\subsection{Long quiver limit}

In the long quiver limit, for $L\to\infty$, the nodes in the quiver diagram in Fig.~\ref{fig:qiuv} become continuously distributed along a circle, effectively building up an additional dimension. This raises the question of whether the four-dimensional quiver theory ${\sf Q}_L$ can be approximated   by a five-dimensional model in this limit, and if so, what its properties are. We argue below that the answer depends crucially on how the 't Hooft coupling constant compares to 
$L$. 

To understand the properties of the long quiver regime, it is convenient to interpret the matrix integral \re{eq:partition function on S4} as defining a one-dimensional periodic lattice model of length $L$.  The nodes of the quiver diagram correspond to the sites of this lattice, each of which possess $N^2-1$ degrees of freedom represented by a $su(n)$ matrix $a_I$. Their dynamics is determined by a nearest-neighbour interaction between $a_I$ and $a_{I\pm 1}$. It is described by the potential \re{eq:interaction potential} and encodes the propagation of excitations along the lattice. Owing to the cyclic symmetry of the interaction potential, these excitations possess a well-defined quasimomentum
\begin{align}\label{quasi}
	p_\alpha={2\pi\alpha\over L}\,,
\end{align}
where $0\le \alpha\le L-1$. 

Within the lattice model, the operators $T_{\alpha,n}(x)$, defined in \re{T-def}, are interpreted as creation operators of $n-$particle excitations with the total quasimomentum $p_\alpha$ and the $U(1)$ charge $n$. The dynamics of these excitations are encoded in the correlation functions  \re{eq:2p intro} in the quiver theory. The selection rule for the parameters $\alpha_i$ that ensures nonvanishing correlators 
is a direct consequence of the quasimomentum conservation.
The relation \re{T-def} provides the decomposition of an excitation of definite quasimomentum \re{quasi}  into a superposition of excitations localized at different sites. 

The correlations between different sites of the lattice model are described by the two-point correlation functions \re{OO} and take the following form  
\begin{align}\label{OO-f}
	\vev{O_n^I(x_1)\bar O_n^J(x_2)} = {\cG_n\over |x_1-x_2|^{2n}}f_{I-J}(g)\,,
\end{align}
where the normalization factor $\cG_n$ is defined in \re{eq:coefficients}. The function $f_{I-J}$ depends on the distance between the two sites $Y=|I-J|$ and it is given by a discrete Fourier transform of ${R}_{\alpha,n}(g)$
\begin{align}\label{f-R}
	f_{Y}=\frac{1}{L}\sum_{\alpha=0}^{L-1} e^{\frac{2\pi i \alpha}{L}Y} R_{\alpha,n}(g) \,.
\end{align}
This function  measures the correlation between two nodes of the quiver gauge theory separated by the distance $Y$. It
depends on the scaling dimension of the operators $n$, the length of the quiver $L$ and it satisfies the periodicity condition
$f_{Y}=f_{Y+L}$. 

As mentioned above, the behaviour of the quiver theory for $L\to\infty$ is determined by how the 't Hooft coupling scales in this limit. The reason for this is that  the coefficients $C_{ij}$ in the interaction potential \re{eq:interaction potential} depend on the 't Hooft coupling as $C_{ij} \sim g^{i+j}$. 
These coefficients govern the interaction strength between adjacent lattice sites, while the interaction between two distant sites $I$ and $J$ scales as the $|I-J|$-th power of the matrix $C$. 

In the weak coupling regime, the interaction strength decreases rapidly with increasing distance between the sites, effectively restricting the interactions to be short-range. In this case, the excitations propagating across the lattice have small values of the quasimomenta \re{quasi} and are described by an effective one-dimensional field theory. Within the context of the quiver theory, this corresponds to the emergence of an additional dimension. 

In contrast, in the strong coupling regime, for $g\gg L\gg 1$, the interaction strength becomes uniform across all site pairs. As a result, the lattice model exhibits strong correlations, and the notion of site ordering along the lattice becomes meaningless. In this case, excitations carry finite quasimomentum \re{quasi}, and the lattice effectively collapses to a point. Consequently, the quiver theory remains a four-dimensional, strongly coupled theory without an emergent dimension.

Based on this analysis, we expect that the quiver theory effectively becomes five-dimensional for $L\to\infty$ provided that the ratio $g/L$ remains finite. In this limit, the quasimomentum \re{quasi} becomes continuous and, as a consequence, the operators \re{T-def} are replaced by their five-dimensional counterparts in the momentum and coordinate representations, respectively,
\begin{align}
	T_{\alpha,n}(x) \  \mapsto \  T_n(x,p)\,,\qqqquad
	O_n^{I} (x) \  \mapsto \  O_n(x,y)\,.
\end{align}
Here $p$ is the momentum of excitations and $y$ is a conjugated five-dimensional coordinate. The two operators $T_n(x,p)$ and $O_n(x,y)$ are related to each other through a Fourier transform. 

Since the correlation functions \re{OO-f} have a factorized dependence on the four-dimensional and quiver coordinates, their five-dimensional counterparts  should have a similar structure, e.g.
\begin{align}\label{eq:two-point with y}
	\vev{O_n(x_1,y_1) \bar O_n(x_2,y_2)} = {\cG_n \over |x_1-x_2|^{2n}}f_n(y_1-y_2)\,.
\end{align}
The function $f_n(y)$ is related to $f_{Y}$ defined in \re{OO-f} in the limit when $Y$, $L$ and $g$ go to infinity with the ratios $y=Y/(2g)$ and $g/L$ held fixed. 
The three-point function has the similar form, see \re{OOOb} below. 

In the following sections, we compute the two-point correlation function \re{eq:two-point with y}  in the limit $L\to\infty$ and identify the corresponding function $f_n(y_1-y_2)$ of the additional $y-$coordinates. As mentioned in the Introduction, the dependence of \re{eq:two-point with y} on five-dimensional coordinates $(x_i,y_i)$  is fixed by the conformal symmetry in four dimensions. Due to the factorized 
dependence of \re{eq:two-point with y} on the $y-$coordinates, an effective five-dimensional theory describing the long quiver limit is not Lorentz invariant.

\section{Weak coupling regime}\label{sect4}

In this section, we compute the two-point correlation function \re{eq:coefficients} at weak coupling and  analyze its behaviour across the different regimes outlined in \re{regimes}. 
 
\subsection{Fredholm determinants at weak coupling}

The dependence of the correlation function \re{eq:coefficients} on the coupling constant resides in the function $R_{\alpha,n}(g)$ defined in \re{eq:twisted two-point}. This function is given by the ratio of Fredholm determinants of the semi-infinite matrices \re{eq:matrix}. 

Changing the integration variable in \re{eq:matrix} as $t\to (2gt)^2$ and replacing the Bessel functions with their expansion around the origin, we observe that the matrix elements scale at weak coupling as $(K_\ell)_{ij}=\cO(g^{2(\ell+i+j-1)})$. This property allows us to expand the determinants in \re{eq:twisted two-point} in powers of the matrices
\begin{align}\label{R-weak}
R_{\alpha,n} = 1- s_\alpha \Big(\tr K_{n+1} -\tr K_{n-1}\Big)+ O(g^{4n})\,,
\end{align}
where the last term describes subleading corrections. It consists of products of traces involving higher powers of the  matrices $K_\ell$, and is suppressed by the factor of $g^{2n}$ relative to the second term. 

Each term of the expansion \re{R-weak} can be computed by employing the explicit expression of the matrix elements (\ref{eq:matrix}). For instance, the leading term in \re{R-weak} takes the following form 
\begin{align} 
	\label{eq:leading single trace}
		\tr K_{\ell} 
		 = g^{2(\ell+1)} \sum_{m\ge 0}^\infty (-1)^m g^{2m} 
		\frac{(2\ell+2m)! q_{\ell+m+1}}{m! [(m+\ell+1)!]^2  (2\ell+m)!}\,,
\end{align} 
where the notation was introduced for the moments of the symbol function defined in \re{eq:symbol}
\begin{equation}
	\label{eq:qn}
		q_n = 2n \int_0^\infty d x \, x^{2n -1}\chi(x) =-4 (2n)!\zeta(2n-1)\,.
\end{equation} 
Note that the expansion coefficients in \re{eq:leading single trace} are proportional to odd Riemann zeta values $\zeta(2n-1)$.

Combining the above relations we obtain from \re{R-weak} the weak coupling expansion 
\begin{equation}
\label{eq:perturbative expansio twisted}
R_{\alpha,n}(g) =1-4g^{2n} s_\alpha  \binom{2n}{n}\zeta(2n-1)+8g^{2n+2}s_\alpha \binom{2n+2}{n+1}n\zeta(2n+1)+ O(g^{2n+4}) \, . 
\end{equation} 
This relation coincides with the result derived in \cite{Billo:2021rdb} by a direct  Feynman diagram calculation of the two-point correlation function $\langle T_{\alpha,n}({x}) {\widebar T}_{\alpha,n}(0) \rangle$ in the quiver theory. In this approach, the $O(g^{2n})$ term in \re{eq:perturbative expansio twisted} arises from a Feynman diagram in which $n$ scalar lines connect the points $x$ and $y$ via an irreducible $n\to n$ transition vertex (see Section 3 of \cite{Billo:2022lrv} for details).

The subleading corrections to \re{eq:perturbative expansio twisted} can be organized according to the power of $s_\alpha$. Specifically, the terms proportional to $s_\alpha^m$ arises from traces of $m$ matrices \re{eq:matrix} and their contribution to \re{eq:perturbative expansio twisted} starts at order $O(g^{2mn})$.  As a result, the expansion (\ref{eq:perturbative expansio twisted}) takes the following form
\begin{align}\label{R-gen}
R_{\alpha,n}(g) =1+ \sum_{m\ge 1} \lr{s_\alpha \, g^{2n}}^m Q_m(g^2)\, , 
\end{align}
where $Q_m(g^2)$ are  series in $g^2$,  with expansion coefficients expressed as multilinear combinations of odd Riemann zeta-values of a homogenous transcendental weight (see \cite{Belitsky:2020qir} for details). Schematically,
\begin{align}\label{Q-gen}
Q_m =\sum_{p\ge 0}  g^{2p} \sum_{\{k\}}  c_{k_1,k_2,\dots,k_m} \zeta(2k_1-1) \zeta(2k_2-1)\dots \zeta(2k_m-1)\,,
\end{align}
where the sum goes over $m$ positive integers $k_1\ge k_2 \ge \dots k_m \ge n$ satisfying the relation
\begin{align}\label{weight}
\sum_{i=1}^m (2k_i-1) = m(2n-1)+2p\,.
\end{align}
The expansion coefficients $c_{k_1,k_2,\dots,k_m}$ are rational numbers depending on $p$.

For $m=1$, it follows from \re{weight} that $k_1=n+p$. Replacing $Q_1(g^2)$ in \re{R-gen} with its general expression \re{Q-gen}, we reproduce \re{eq:perturbative expansio twisted} and identify the values of the coefficients
\begin{align}\label{qn}
c_n= -4 \binom{2n}{n}\,,\qqqquad c_{n+1}=8 n\binom{2n+2}{n+1}\,,\qqquad \dots
\end{align}
For arbitrary $m\ge 2$, the leading $O(g^0)$ term in \re{Q-gen} has $k_1=\dots=k_m=2n-1$. The corresponding coefficient takes a remarkably simple form $c_{n,\dots,n} = c_n^m$ leading to 
\begin{align}
Q_m=[c_n\zeta(2n-1)]^m +O(g^2)\,,
\end{align}
where $c_n$ is given by (\ref{eq:qn}). Substituting this relation into \re{R-gen} and neglecting $O(g^2)$ corrections to $Q_m$, we finally find
\begin{align}\label{R-app}
R_{\alpha,n}(g) ={1\over 1-s_\alpha g^{2n} c_n \zeta(2n-1)} + \dots \,.
\end{align}
This relation holds for sufficiently small values of the coupling constant. It can be derived directly from \re{eq:twisted two-point} by noticing that 
${(K_{n-1}^m)}_{11}=[{(K_{n-1})}_{11}]^m(1+O(g^2))$.~\footnote{The relation \re{R-app} can also be derived from the first equation in \re{eq:twisted two-point} by expanding $R_{\alpha,n}(g)$ over the eigenvalues of the matrix $K_{n-1}$. The relation \re{R-app} captures the contribution of the largest eigenvalue.}
In the perturbative calculation of the correlation function $\langle T_{\alpha,n}({x}) {\widebar T}_{\alpha,n}(0) \rangle$, the relation \re{R-app} arises from Feynman diagrams in which $n$ scalar lines connect the points  $x$ and $0$ and pass through an arbitrary number of consecutive $n \to n$ transition vertices. The sum of these diagrams forms a geometric progression.

\subsection{Convergence properties}

Before discussing the correlation functions, let us consider the convergence properties of the weak coupling expansion \re{R-gen}.
By examining the ratio of successive coefficients in the expansion of \re{R-gen} in powers of $g^2$, it is straightforward to verify that, for arbitrary $n$ and $\alpha\neq 0$, this ratio approaches a finite value $(-16)$. This suggests that the weak coupling expansion \re{R-gen} has a finite radius of convergence $g_\star^2=-1/16$.

To demonstrate this, we replace the moments $q_n$  in \re{eq:leading single trace} with their leading large $n$ behaviour $q_n\sim -4 (2n)!$. The resulting expression for $\tr K_\ell$ develops a logarithmic singularity at $g^2=-1/16$. 
Moreover, introducing $\delta=1+16g^2$, we find that the leading term in \re{R-weak} takes the following form for $\delta\to 0$
\begin{align}\label{R-LO}
R_{\alpha,n}(g)= -  n r \delta \log \delta + \dots\,,
\end{align}
where $r =4(-1)^ns_\alpha/\pi$ and dots denote terms analytical in $\delta$ (including the finite terms), as well as subleading  contributions resulting from  \re{R-weak}.

The subleading corrections to \re{R-LO} can be obtained by expanding the first relation in \re{eq:twisted two-point} in powers of the $K-$matrix
\begin{align}\label{R-NLO}
 {R}_{\alpha,n}(g) = 1+s_\alpha {(K_{n-1})}_{11} +s^2_\alpha {(K^2_{n-1})}_{11} +\dots
\end{align}
The first two terms in this relation give rise to \re{R-LO}, while the contributions of the remaining terms in \re{R-NLO} can be analyzed using the techniques described in \cite{Bajnok:2024ymr}. By carrying out this calculation, we find that for $\delta\to 0$
\begin{align}\label{logs}
s^m_\alpha {(K^m_{n-1})}_{11} =  - n \, \delta \lr{{r^m\over m!}  \log^m \delta + O(\log^{m-1} \delta )} + \dots\,,
\end{align}
where $r$ is defined in \re{R-LO} and dots denote terms analytical in $\delta$.  

Combining the above relations we obtain the leading behaviour of ${R}_{\alpha,n}(g)$ for $g\to g_\star$
\begin{align}
 {R}_{\alpha,n}(g) &=  {R}_{\alpha,n}(g_\star)-n \, \delta^{1+4(-1)^ns_\alpha/\pi} + \dots\,,
\end{align}
where $\delta=1+16g^2\to 0$.
We would like to emphasize that this relation was derived by neglecting   subleading corrections to \re{logs}. Strictly speaking, it is valid in the limit $\delta\to 0$ while keeping the product $s_\alpha \log \delta$ fixed.

\subsection{Correlation functions at weak coupling}

We can apply the relations  \re{eq:perturbative expansio twisted} and \re{R-gen} to derive the weak coupling expansion of the function $f_Y$ defined 
in \re{OO-f} and \re{f-R}. The first two terms of the expansion look as
\begin{align}\label{fY-0}
f_Y=\delta_{Y,0} + \lr{\delta_{Y,1}-2\delta_{Y,0}+\delta_{Y,-1}} g^{2n} \binom{2n}{n}\zeta(2n-1) + O(g^{2n+2})\,,
\end{align}
where the Kronecker delta-function $\delta_{Y,p}$ imposes the condition $Y=p$ (mod $L$) on the lattice. 
From a physical point of view, the three terms inside the brackets in the second term
in \re{fY-0} describe the hopping of excitations to the adjacent sites on the lattice.

In the similar manner, the terms inside the sum in \re{R-gen} describe the hopping of excitations to the adjacent $m$ sites. 
Their contribution to $f_Y$ is given by
\begin{align}\label{fY}
f_Y=\delta_{Y,0} +\sum_{m\ge 1} g^{2nm}  f_{m,Y} Q_m(g^2)\, ,
\end{align}
 where the function $f_{m,Y} $ arises from the Fourier transform of the function $s^m_\alpha$  defined in \re{eq:quantities}
\begin{align}\label{waves}
f_{m,Y} ={1\over L}\sum_{\alpha=0}^{L-1}e^{{2\pi i\over L}Y\alpha}s_\alpha^m\,. 
\end{align}
The correlation  function \re{fY} satisfies $f_{Y}=f_{-Y}=f_{L-Y}$, allowing us to restrict $Y$ to the interval $0\le Y\le L$.

For $m < L/2$, the function \re{waves} is given by linear combinations of Kronecker delta functions $\delta_{Y,\pm p}$ with $0 \le p \le m$. As a consequence, $f_{m,Y}$
vanishes for $m<\min(Y,L-Y)$ and the sum in \re{fY} starts at $m=\min(Y,L-Y)$.
For $m \ge L/2$, the partial waves \re{waves} are different from zero for any $Y$ and they take a more complicated form. This is because, starting from $m = L/2$, the interaction range $2m$ exceeds the length of the lattice. As a result, for $m+Y> L$ the excitations can wrap the lattice, leading to the additional finite-size corrections to $f_{m,Y}$.

The resulting expression  for \re{waves} looks as
\begin{align}\label{f-gen}
f_{m,Y}=\sum_{\ell=-\infty}^{\infty}  {(-1)^{Y-\ell L}\,\Gamma(2m+1)\over 4^m \Gamma(m+Y-\ell L +1)\Gamma(m-Y+\ell L +1)}\,.
\end{align}
This relation holds for arbitrary integer $Y$ and nonnegative integer $m$. The sum over $\ell$ ensures that $f_{m,Y}$ is periodic on the lattice,  $f_{m,Y}=f_{m,Y+L}$, and satisfies the relations $f_{m,Y}=f_{m,-Y}=f_{m,L-Y}$. The $\Gamma-$functions in the denominator of \re{f-gen} effectively truncate the infinite sum in \re{f-gen} to $ (Y-m+1)/L< \ell < (Y+m+1)/L$.
For $0\le m <L/2$  the sum in \re{f-gen} contains only one term. For $L/2\le m<L$ there are at most one additional  term, for $L\le m< 3L/2$ there are at most two additional terms, etc. These additional terms describe the wrapping corrections to \re{f-gen}. 

Combining together \re{fY} and \re{f-gen} and replacing the functions $Q_m(g^2)$ with their perturbative expansion \re{Q-gen}, we can compute the function $f_Y$ at weak coupling for arbitrary number of nodes, $L$, and the scaling dimension of the operators, $n$. 
The leading behaviour of $f_Y$ at weak coupling comes from the first term in the sum in \re{fY}. For $0< Y\le L/2$
we find 
\begin{align}\label{fY-LO}
f_Y=   g^{2nY}  f_{Y,Y} Q_Y(g^2)+ \dots =\left[{g^{2n} \zeta(2n-1)\binom{2n}{n}}\right]^Y+ \dots\,.
\end{align}
For $Y>L/2$ we have instead $f_{Y}=f_{L-Y}$.

It follows from \re{fY-LO} that, at weak coupling, the correlation function decays exponentially fast with the separation $Y$ between the two sites 
\begin{align}\label{mu-weak} 
f_Y = e^{-\mu_n Y} \,,\qqqquad \mu_n = -\log\left({g^{2n} \zeta(2n-1)\binom{2n}{n}}\right)\,.
\end{align}
This relation holds for arbitrary $0< Y\le L/2$ up to corrections suppressed by powers of $g^2$.
The parameter $\mu_n$ has the meaning of a `mass scale' (or inverse correlation length) at weak coupling. 

\begin{figure}[t!]
	\begin{center}
				\includegraphics[width=0.65\textwidth]{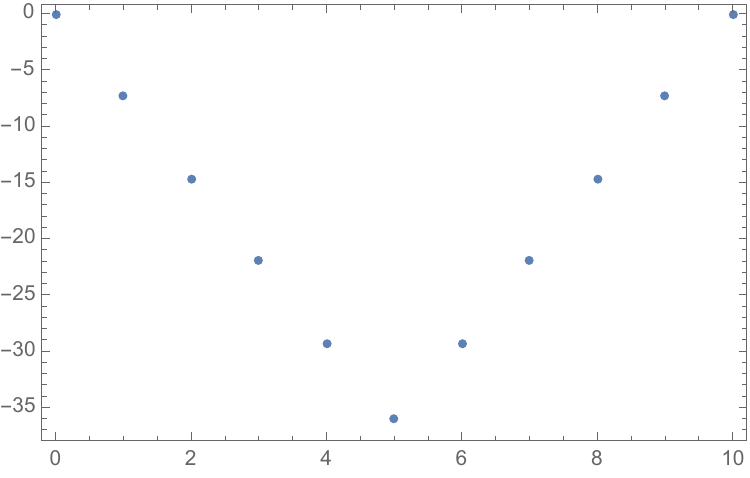}
		\caption{Dependence of the correlation function \re{OO-f} on the distance between the sites $Y=|I-J|$ for
		$n=2$, $g=0.1$ and $L=10$. The horizontal and vertical axis represent $Y$ and $\log f_Y$, respectively. The correlation function is symmetric about $Y=L/2$.}
		\label{fY-plot}
	\end{center}
\end{figure}

To test the relation \re{mu-weak}, we put $n=2$ and $L=10$, expanded the first relation in \re{eq:twisted two-point} at weak coupling up to order $O(g^{40})$ and used \re{R-gen} to compute the functions $Q_m(g^2)$ for $0\le m\le 10$. Substituting these functions into \re{fY}, we computed $f_Y$ for $g=0.1$ and integer $Y$'s within the interval $0\le Y\le 10$. The results are shown in Fig.~\ref{fY-plot}. We observe an agreement with 
\re{mu-weak}.

\subsection{The long quiver limit at weak coupling}

The relation \re{fY-LO} accounts for the leading correction to $f_Y$ at weak coupling.
Higher-order corrections to \re{fY-LO} do not alter the exponential behaviour described in \re{mu-weak}. 
To show this, we apply \re{R-app} and calculate its discrete Fourier transform \re{f-R}~\footnote{We use the symmetry of the sum in \re{f-R} under $\alpha\to L-\alpha$ to change the range of $\alpha$.}
\begin{align}\label{fY-sum}
f_Y =\frac{1}{L}\sum_{\alpha=-L/2}^{L/2-1}  {e^{\frac{2\pi i \alpha}{L}Y}\over 1+4\sin^2\left(\frac{\pi\alpha}{L}\right) {e}^{-\mu_n}} \,,
\end{align}
where $\mu_n$ is defined in \re{mu-weak}.
Remarkably, this sum admits a closed-form evaluation, yielding
\begin{align}\label{eq:improved_f_y}
f_Y &=  {e}^{-\mu_n Y} \frac{\left(\frac{2}{1+\sqrt{1+4  {e}^{-\mu_n}}}\right)^{2Y}}{\sqrt{1+4  {e}^{-\mu_n}}\left[1-\left(\frac{2 {e}^{-\mu_n}}{1+\sqrt{1+4  {e}^{-\mu_n}}+2  {e}^{-\mu_n}}\right)^L\right]} + (Y\to L-Y) \,,  
\end{align}
where $\mu_n$ is defined in \re{mu-weak}. This relation is valid for $0\le Y\le L$ and arbitrary $n$ and $L$. We would like to emphasize that it was obtained by neglecting $O(g^2)$ corrections in \re{R-app}.
The second term in \re{eq:improved_f_y} ensures the invariance of $f_Y$ under $Y \to L-Y$. The two terms in \re{eq:improved_f_y} 
correspond to the two distinct propagation paths of excitations between sites $I$ and $J$, separated by a distance $Y = |I-J|$, across the quiver diagram shown in Fig.~\ref{fig:qiuv}. 

At weak coupling, the mass scale in \re{mu-weak} becomes large and the relation \re{eq:improved_f_y} can be expanded in powers of $e^{-\mu_n}$
\begin{align}\label{fY-exp} 
f_Y{}&=e^{-\mu_n Y}\left[ 1-2(1+Y)e^{-\mu_n} +(2+Y)(3+2Y) e^{-2\mu_n} 
+O(e^{-3\mu_n})\right] +(Y\to L-Y)\,.
\end{align}
The first term dominates for $0<Y<L/2$ and the second one for $L/2<Y<L$. The turnover occurs at $Y=L/2$. The first term in \re{fY-exp} differs from \re{mu-weak} only by the expression inside the brackets. It accounts for subleading corrections to $f_Y$ and is given by a series in $e^{-\mu_n}=O(g^{2n})$. Since the coefficients of this series are polynomials in $Y$,  the asymptotic behaviour of $f_Y$ at large $Y$ is unchanged.  

The dependence of \re{fY-exp} on the scaling dimension of the operators, $n$, enters through $\mu_n$, see \re{mu-weak}. At fixed coupling and large $n$, this parameter scales as $\mu_n\sim n \log(1/(4g^2))$ and grows linearly with $n$. For fixed $Y\neq 0$ the function $f_Y$ vanishes exponentially fast at large $n$ leading to 
\begin{align}
\lim_{n\to\infty} f_Y=\delta_{Y,0}\,.
\end{align}
This relation also follows immediately from \re{fY-0}.

The preceding expressions hold for quivers of arbitrary length $L$. Let us now investigate their behaviour for $L \to \infty$.
As we already noted, the quasimomentum \re{quasi} becomes continuous in this limit,  $p_\alpha \to p$,  and takes values within the first Brillouin zone $[-\pi,\pi)$. This suggests that the sum \re{fY-sum} can be approximated by an integral over $p$. Consequently, in the continuum limit, for $L\to\infty$, and for fixed  $Y$  the relation \re{fY-sum} reduces to  
\begin{align}\label{fY-cont} 
\lim_{L\to\infty} f_Y=\int_{-\pi}^{\pi}\dfrac{d p}{2\pi}{e^{i p Y}\over 1+4\sin^2(p/2) e^{-\mu_n}} = \mathrm{e}^{-\mu_n Y} \frac{\left(\frac{2}{1+\sqrt{1+4{e}^{-\mu_n}}}\right)^{2Y}}{\sqrt{1+4{e}^{-\mu_n}}}\,.
\end{align}
We verify that this relation is in a perfect agreement with \re{eq:improved_f_y}.

We conclude that, in the weak coupling regime, the correlations between different sites in the quiver diagram in Fig.~\ref{fig:qiuv} decrease exponentially with the separation between the sites $Y$. The corresponding mass scale \re{mu-weak} depends on the scaling dimension of the operators $\Delta=n$ and scales logarithmically with the coupling constant, $\mu_n\sim n\log(1/(4g^2))$.

\section{Strong coupling regime}
\label{sec:Sec5}

In the strong coupling regime, it is convenient to use the representation of the correlation functions \re{C123-GF} in terms of the Fredholm determinant of the Bessel kernel \re{F}. We summarize below the properties of these determinants and, then, compute the correlation functions in the different regimes \re{regimes}. 

\subsection{Strong coupling expansion of the Fredholm determinant}

For arbitrary values of the parameters, the strong coupling expansion of the function $\mathcal F_\ell$ defined in \re{F} can be obtained by employing  the results of \cite{Beccaria:2022ypy,Beccaria:2023kbl,Bajnok:2024epf,Bajnok:2024ymr}. 
This expansion has the following general form 
\begin{align}\label{F-exp}
\mathcal F_\ell = 4\pi g \,a(1-a) -\frac12(2\ell-1)\log g + B_\ell + \Delta \mathcal F_\ell(g)\,, 
\end{align}
where the notation was introduced for $a=\alpha/L$. 

Each successive term on the right-hand side of \re{F-exp} is smaller than the preceding one. The constant term $B_\ell$ depends on $a$ and 
satisfies
\begin{align}
B_{\ell+1} -B_\ell =\log \ell-\frac12 \log(4s_\alpha) \,.
\end{align} 
Its explicit expression can be found in \cite{Beccaria:2023qnu}.

The function $\Delta \mathcal F_\ell(g)$ vanishes as $g\to\infty$. It accounts for subleading terms in \re{F-exp} of two different types: `perturbative' corrections, proportional to powers of $1/g$, and `nonperturbative', exponentially small corrections that are proportional to 
\begin{align}\label{Lambdas}
\Lambda^2_-=e^{-8\pi ga} \,,\qqqquad \Lambda^2_+ =e^{-8\pi g(1-a)} \,.
\end{align}
These two nonperturbative parameters are related by the transformation $\alpha\to L-\alpha$ and satisfy the relation $\Lambda^2_-\Lambda^2_+=e^{-8\pi g}$. 
The resulting expression for $\Delta \mathcal F_\ell(g)$ takes the form of a transseries
\begin{align}\label{FFs}
\Delta \mathcal F_\ell(g)= \mathcal F_\ell^{(0,0)}(g) + \sum_{n,m} \Lambda_-^{2n} \Lambda_+^{2m} \mathcal F_\ell^{(n,m)}(g)\,,
\end{align}
where the coefficient functions $\mathcal F_\ell^{(0,0)}(g)$ and $\mathcal F_\ell^{(n,m)}(g)$ are asymptotic series in $1/g$. Due to 
the symmetry of $\Delta \mathcal F_\ell(g)$ under $\alpha\to L-\alpha$, the functions  $\mathcal F_\ell^{(n,m)}$ and $\mathcal F_\ell^{(m,n)}$ are related by the same transformation.

The first term on the right-hand side of \re{FFs} encodes the perturbative corrections at strong coupling
\begin{align}\notag\label{F00}
\mathcal F_\ell^{(0,0)}={}&-{1\over 16g}(2\ell-1)(2\ell-3)I_1 -{1\over 64g^2}(2\ell-1)(2\ell-3) I_1^2
\\\notag
{}& -{1\over 3072 g^3} (2\ell-1)(2\ell-3)[(2\ell-5)(2\ell+1)I_2+16 I_1^3] 
\\
{}& 
-\frac{1}{2048 g^4}(2\ell-1)(2\ell-3)\left[(2 \ell-5) (2\ell+1)I_1 I_2+4 I_1^4\right]+O(1/g^5)\,.
\end{align}
Expressions for the subleading terms in this relation can be found in \cite{Belitsky:2020qir}. To all orders in $1/g$, the expansion coefficients in \re{F00} are proportional to $(2\ell-1)(2\ell-3)$. They are given by multilinear combinations of the functions $I_n=I_n(a)$ defined as
\begin{align}\label{In}
I_n  = \dfrac{(-1)^n}{(2\pi)^{2n-1}(2n-2)!}\left[\psi^{(2n-2)}\left(a\right)+\psi^{(2n-2)}\left(1-a\right)-2\psi^{(2n-2)}(1)\right],
\end{align}
where  $\psi^{(n)}(x)=\left(d/d x\right)^n \psi(x)$ is a derivative of Euler function $\psi(x)=(\log \Gamma(x))^\prime$. Note that the expansion coefficients in \re{F00} involve powers of $I_1$. All such terms depending on this function can be eliminated through the redefinition of the coupling constant 
\begin{align}\label{g1}
g'=g-\frac12 I_1(a)\,,\qqqquad a=\alpha/L\,.
\end{align}

Let us stress that the expansion coefficients in \re{F00} grow factorially at large orders \cite{Beccaria:2022ypy,Bajnok:2024ymr,Bajnok:2024epf}, rendering the series divergent and necessitating regularization. Nonperturbative terms in \re{FFs} are needed to compensate the inherent ambiguities arising from the choice of Borel resummation of $\mathcal{F}_\ell^{(0,0)}(g)$ and $\mathcal{F}_\ell^{(n,m)}(g)$. These functions are related to each other by resurgence relations \cite{Marino:2012zq,Dorigoni:2014hea,Aniceto:2018bis}.

The leading nonperturbative function in \re{FFs} is given by
\begin{align}\notag\label{F10} 
\mathcal F_\ell^{(1,0)}={}& -i(-1)^\ell\mathcal S(a) 
 \bigg[1+\frac{(2 \ell-3) (2 \ell-1)}{16\pi  a g'} +\frac{(2 \ell-5) (2 \ell-3)  (2 \ell-1) (2 \ell+1)}{512 (\pi a g')^2} 
\\[2mm] \notag
{}&\quad + \frac{(2 \ell -5) (2 \ell -3) (2 \ell -1) (2 \ell +1) ((2 \ell -7) (2 \ell +3) +6)}{24576 (\pi a g')^3}
\\[2mm]
{}&\quad + \frac{(2 \ell -5) (2 \ell -3) (2 \ell -1) (2 \ell +1) \left((2 \ell -7)^2 (2 \ell +3)^2+768 (\pi a)^3 I_2\right)}{1572864 (\pi a g')^4} +O(1/g^5)\bigg], 
\end{align}
where the coupling $g'=g'(a)$ is defined in \re{g1} and $\mathcal S(a)$ is the Stokes constant
\begin{align}\label{Stokes}
\mathcal S(a)=\frac{\Gamma (1-2 a) \Gamma^2 (1+a)}{\Gamma (1+2 a)\Gamma^2 (1-a)} \,.
\end{align}
The function $\mathcal F_\ell^{(0,1)}$ can be obtained from \re{F10} by replacing $a\to 1-a$.  
Expansion of \re{F10} in powers of $1/g$ generates terms proportional to powers of $I_1$.  Expressions for the subleading corrections to \re{F10} can be found in \cite{Bajnok:2024epf,Bajnok:2024ymr}.  

\subsection{The correlation function at strong coupling} 

Using the expressions for the strong coupling expansion of the function $\mathcal F_{\alpha,n}$, we can apply the relations \re{eq:toda equation} to get the analogous expressions for the functions $R_{\alpha,n}$ and ${\mathcal{V}}_{\alpha,n}$
\begin{align}\notag\label{RV}
{}& R_{\alpha,n}={(n-1)n\over 4g^2 s_\alpha} \times  R^{(0,0)} \left[1 + e^{-8\pi ga} \, R^{(1,0)} +e^{-8\pi g(1-a)}\, R^{(0,1)} +\dots \right] ,
\\[2mm]
{}& {\mathcal{V}}_{\alpha,n} =\sqrt{n-1\over n} \times \mathcal{V}^{(0,0)} \left[1 + e^{-8\pi ga}\, \mathcal{V}^{(1,0)} + e^{-8\pi g(1-a)}\, \mathcal{V}^{(0,1)} +\dots \right].
\end{align}
In both relations, the first factor yields the leading asymptotic behaviour at large $g$, while dots denote subleading corrections suppressed by powers of the nonperturbative parameters $\Lambda_-^2$ and $\Lambda_+^2$ defined in \re{Lambdas}. 

The perturbative functions $R^{(0,0)}$ and $\mathcal{V}^{(0,0)}$ are related to the function $\mathcal F_\ell^{(0,0)}$ defined in \re{F00}  by the same relations as \re{eq:toda equation}. In a close analogy with \re{F00}, they are given by an asymptotic series in $1/g$ with coefficients that are polynomial in $I_n(a)$. To eliminate the dependence on $I_1(a)$, we expand $R^{(0,0)}$ and $\mathcal{V}^{(0,0)}$ in inverse powers of the coupling \re{g1}
\begin{align}\notag\label{RV00}
{}&  R^{(0,0)} = \lr{g'\over g}^{2(n-1)} \bigg[1-\frac{(n-1) (2 n-3) (2 n-1)}{96 {g'}^3}I_2(a) 
 \\\notag
{}&\hspace*{35mm} 
-\frac{3 (n-1) (2 n-5) (2 n-3) (2 n-1) (2 n+1)}{10240 {g'}^5}I_3(a) +O(1/g'{}^6) \bigg],
\\
\notag
{}& \mathcal{V}^{(0,0)}=  \lr{g\over g'}^{1/2}\bigg[1+\frac{(2 n-3) (2 n-1)}{128 {g'}^3}I_2(a)
 \\
{}&\hspace*{30mm} 
 +\frac{3(2 n-5) (2 n-3) (2 n-1) (2
   n+1)}{8192 {g'}^5} I_3(a) +O(1/{g'}^6) \bigg],
\end{align}
where the coupling $g'=g'(a)$ is defined in \re{g1}.
Note that the expansion coefficients of the series within the brackets are polynomials in $n$ with a degree that matches the power of $1/g'$ for $R^{(0,0)}$, while is one degree lower for $\mathcal{V}^{(0,0)}$. 

As follows from \re{eq:toda equation}, the nonperturbative functions in \re{RV} are related to the function \re{F10} as $R^{(1,0)} =\mathcal F_{n+1}^{(1,0)}-\mathcal F_{n-1}^{(1,0)}$ and 
$\mathcal{V}^{(1,0)} = \mathcal F_{n}^{(1,0)}-(\mathcal F_{n+1}^{(1,0)}+\mathcal F_{n-1}^{(1,0)})/2$. Their expressions are 
\begin{align}\notag\label{RV10}
{}& R^{(1,0)} = i (-1)^n \mathcal S(a) \frac{(n-1)}{\pi  a g'}\bigg[1+\frac{(2 n-3) (2 n-1)}{16 \pi  a {g'}}
\\\notag
{}& \hspace*{50mm} +\frac{(2 n-3) (2 n-1)
   \left(4 n^2-8 n-1\right)}{512 (\pi  a {g'})^2}+O(1/g^3)\bigg],
\\ \notag
{}& \mathcal{V}^{(1,0)} =-2 i (-1)^n \mathcal S(a)\bigg[1+\frac{4n^2-8 n+5}{16 \pi  a g'}
\\
{}&\hspace*{42mm} 
+\frac{(2 n-3) (2 n-1) \left(4 n^2-8 n+7\right)}{512 (\pi a g')^2}+O(1/g^3)\bigg],
\end{align}
where  $\mathcal S(a)$ is given by \re{Stokes}. Let us note that the expansion coefficients in \re{RV10} are polynomials in $n$ with a  degree that is twice the power of $1/g'$. 
The functions $R^{(0,1)}$ and $\mathcal{V}^{(0,1)}$ can be obtained from \re{RV10} by replacing $a\to 1-a$. Expressions for the subleading corrections in \re{RV} can be found in \cite{Bajnok:2024ymr}.

For finite $a=\alpha/L$, the relations \re{RV} can be used to describe the functions $R_{\alpha,n}$ and ${\mathcal{V}}_{\alpha,n}$ over a wide range of values of the coupling constant, including sufficiently small values within the radius of convergence of the weak coupling expansion.  
While nonperturbative corrections to \re{RV} are exponentially suppressed at large $g$, they become significant in the transition region where 
$g=O(1)$.  

The situation changes significantly for small $a$, or equivalently, for large quiver length $L$. First, the expansion coefficients in the perturbative functions \re{RV00} 
develop poles in $1/a$ as $a \to 0$. Second, the nonperturbative parameter $\Lambda_-^2 = e^{-8\pi ga}$ approaches $1$ in the limit $a \to 0$. Consequently, the nonperturbative corrections to \re{RV}, which are proportional to powers of $\Lambda_-^2$, become indistinguishable from the perturbative corrections. Moreover, the corresponding nonperturbative functions \re{RV10} develop $1/a$ poles, similar to their perturbative counterparts in \re{RV00}. Therefore, determining the functions $R_{\alpha,n}$ and $\mathcal{V}_{\alpha,n}$ in the long-quiver limit $L \to \infty$, or equivalently $a \to 0$, necessitates the resummation of an infinite series of nonperturbative contributions in \re{RV}.

\subsection{The long quiver limit}  

The relations \re{RV} are valid at strong coupling for arbitrary $a=\alpha/L$ and $n$. Let us examine their behaviour in the limit of long quiver $L\to\infty$, or equivalently $a\to 0$. As follows from \re{In}, the functions $I_n(a)$ diverge in this limit as 
\begin{align}\label{I-as}
I_n(a) = {(-1)^{n+1}\over  \lr{2\pi a}^{2 n-1}} + O\lr{a^2}\,.
\end{align} 
Substituting this relation into \re{RV00} and \re{RV10}, we find that, in the long quiver limit, the strong coupling expansion effective runs in powers of the ratio $1/(ag)$. 

This suggests to consider a double scaling limit when both the coupling constant and the length of the quiver go to infinity while their ration is kept fixed,
\begin{align}\label{xi}
L\to\infty \,,\qqqquad g\to\infty\,,\qqqquad \xi={2\pi \alpha g\over L} = \text{fixed}\,.
\end{align}
As a nontrivial test, we verify that the expressions in \re{RV} approach a finite value in this limit. In particular, the function $R_{\alpha,n}$ simplifies as
\begin{align}\label{R-simp}
\widetilde R_{\alpha,n}=(n-1) n\,\xi ^{-2} \,  \widetilde R^{(0,0)} \left[1 + e^{-2\xi} \,\widetilde  R^{(1,0)} +O(e^{-4\xi}) \right],
\end{align}
where we introduced a tilde to indicate that this relation only holds in the limit \re{xi}. 

The nonperturbative corrections in \re{R-simp} run in powers of $\Lambda_-^2=e^{-2\xi}$, the second nonperturbative parameter $\Lambda_+^2=e^{-8\pi g(1-a)}$ vanishes for $g\to\infty$. The coefficient functions in \re{R-simp} are given by series in $1/\xi$
\begin{align}\notag\label{small-xi}
\widetilde R^{(0,0)} {}& =  1-{(n-1)\over \xi} +\frac{1}{4\xi^2} (n-1) (2 n-3)-\frac{1}{32\xi^3} (n-1) (2 n-5) (2 n-3)+O(1/\xi^4)\,,
\\[2mm]
\widetilde R^{(1,0)} {}&=2 i (-1)^n (n-1)\bigg[{1\over\xi}+\frac{1}{8\xi^2 } (4n^2-8n+7) +\frac{1}{128\xi^3} (4n^2-8n+7) (4n^2-8n+11) +O\left(1/\xi^4\right)\bigg],
\end{align}
where $\xi$ is defined in \re{xi}.

It follows from \re{R-simp} and \re{small-xi} that the function $\widetilde R_{\alpha,n}$ has a  different behaviour at small and large $\xi$. 
At large $\xi$, the nonperturbative corrections in \re{R-simp} are exponentially suppressed and the resulting expansion of $R_{\alpha,n}$ in powers of $1/\xi$ is in one-to-one correspondence with the conventional $1/g$ expansion in the AdS/CFT correspodence. 
At small $\xi$, all terms inside the brackets in \re{R-simp} become equally important. As mentioned earlier, finding $R_{\alpha,n}$ in this regime requires resummation of an infinite series of nonperturbative corrections.

Remarkably, the functions \re{RV} can be found in the limit \re{xi} in a closed form for arbitrary $\xi$. A key observation is that, in the dual lattice model description of the quiver theory, the limit \re{xi} corresponds to the vanishing quasimomentum \re{quasi} of the excitations propagating across the quiver diagram. As was shown in \cite{Beccaria:2023qnu}, describing the contribution of these excitations to \re{eq:twisted two-point}, we can simplify the integral representation of the matrix \re{eq:matrix} by replacing the symbol function $\chi\left({\sqrt{t}}/{(2g)}\right)$ by its leading behaviour at large $g$
\begin{align}
s_\alpha \chi\left(\frac{\sqrt{t}}{2g}\right) \ \to\  -\dfrac{(2\xi)^2}{t}\,,
\end{align}
where we used \re{eq:symbol} and \re{eq:quantities} and took the limit \re{xi}. Applying this transformation to \re{eq:matrix}, the matrix $K_\ell$ becomes tridiagonal and its Fredholm determinant \re{F} can be computed exactly (see \cite{Beccaria:2023qnu})
\begin{align}\label{F-limit}
\widetilde{\mathcal F}_\ell = \log \lr{\Gamma(\ell)\,\xi^{1-\ell} \, {\mathrm I}_{\ell-1}(2\xi)}\,,
\end{align}
where ${\mathrm I}_{\ell-1}(2\xi)$  is a modified Bessel function of the first kind.  

\begin{figure}[t!]
	\begin{center}
				\includegraphics[width=0.65\textwidth]{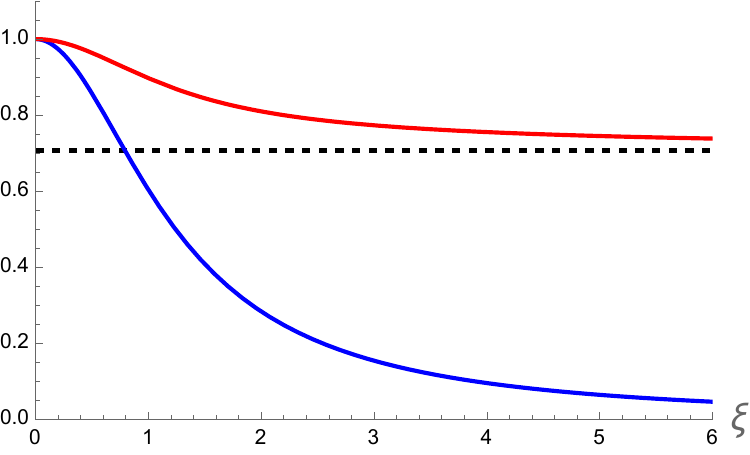}
		\caption{The dependence of the functions $\widetilde  R_{\alpha,n}$ (blue curve) and $\widetilde{\mathcal V}_{\alpha,n}$ (red curve)
		defined in \re{RV-limit} on $\xi$ for $n=2$. The dashed line indicates the limiting value of $\widetilde{\mathcal V}_{\alpha,n}$ for $\xi\to\infty$. }
		\label{RV-plot}
	\end{center}
\end{figure}

Substituting \re{F-limit} into \re{eq:toda equation} we arrive at  
\begin{align}\notag\label{RV-limit}
{}&\widetilde  R_{\alpha,n}=(n-1) n \,\frac{ {\mathrm I}_n(2\xi)}{\xi ^{2} {\mathrm I}_{n-2}(2\xi)}\,,
\\
{}&\widetilde{\mathcal V}_{\alpha,n}=\sqrt{n-1\over n}\left[\frac{{\mathrm I}^2_{n-1}(2\xi)}{{\mathrm I}_{n-2}(2\xi) {\mathrm I}_n(2\xi)}\right]^{1/2}.
\end{align}
We would like to emphasize that these relations hold in the double scaling limit \re{xi} for arbitrary $\xi$. 
The dependence of the functions \re{RV-limit} on $\xi$ is shown in Fig.~\ref{RV-plot}. At large $\xi$, they behave differently, $\widetilde  R_{\alpha,n}\sim (n-1) n/\xi ^2$ and $\widetilde{\mathcal V}_{\alpha,n}\sim \sqrt{(n-1)/n}$, while for small  $\xi$ they approach the same value.

Replacing the Bessel functions in \re{RV-limit} by their asymptotic behaviour at infinity, we verified that expansion of $\widetilde R_{\alpha,n}$ at large $\xi$ is in agreement with \re{R-simp} and \re{small-xi}. In the similar manner, the large-$\xi$ expansion of $\widetilde {\mathcal V}_{\alpha,n}$ coincides with the expansion of the function ${\mathcal V}_{\alpha,n}$ given by \re{RV} in the limit \re{xi}. 

At small $\xi$, or equivalently for $L\gg g$, we find from \re{RV-limit}
\begin{align}\notag\label{RV-small-xi}
{}&\widetilde  R_{\alpha,n}=1-\frac{2\xi ^2}{(n-1) (n+1)}+\frac{(5 n+1)\xi^4}{(n-1)^2 n (n+1) (n+2)}+O(\xi^6)\,,
\\
{}&\widetilde{\mathcal V}_{\alpha,n}=1-\frac{\xi ^2}{(n-1) n (n+1)}+\frac{3 (2 n+1) \xi ^4}{2 (n-1)^2 n (n+1)^2
   (n+2)}+O(\xi^6)\,. 
\end{align}
It is interesting to note that, although this relation was derived in the strong coupling regime, the asymptotic behavior of the functions $\widetilde  R_{\alpha,n}$ and $\widetilde{\mathcal V}_{\alpha,n}$ as $\xi \to 0$ closely resembles that observed in the weak coupling regime (see \re{eq:perturbative expansio twisted}). 

This universality arises because, in both regimes, the coupling constant is much smaller than the quiver length. Consequently, in the long-quiver limit, for $L \gg g$, the correlation functions \re{OO-f} are expected to exhibit the same exponential behavior \re{mu-weak} with respect to the distance between sites, regardless the value of the coupling constant. We explicitly demonstrate this in the next section and calculate the mass scale $\mu_n(g)$ in the strong coupling regime.
 
\subsection{Heavy operators limit}
 
Let us examine the relations \re{RV} in the limit of heavy operators $n\gg 1$, associated with  the last regime in \re{regimes}.
  
We observe from \re{RV00} that, at large $n$, the strong coupling expansion of the perturbative functions $R^{(0,0)}$ and $\mathcal V^{(0,0)}$ runs in powers of $n/g$. Similar to the previous case (see \re{xi}), this suggests to consider the double scaling limit 
\begin{align}\label{large-n}
n\to\infty \,,\qqqquad g\to\infty\,,\qqqquad \hat n={n\over 2g}= \text{fixed}\,.
\end{align}
It follows from \re{RV} that for finite $a=\alpha/L$ the nonperturbative corrections vanish in this limit and the functions ${R}_{\alpha,n}$ and ${\mathcal{V}}_{\alpha,n}$ only receive perturbative contributions.
In particular, we use \re{RV} and \re{RV00} to obtain in the limit \re{large-n}
\begin{align}\label{bar-R}
\log \hat {R}_{\alpha,n} = 2\log \hat n -\log s_\alpha -2 \hat n  I_1(a)-\frac{1}{3}  \hat n^3 I_2(a) +O(\hat n^5)\,,
\end{align}
where we introduced a hat on the left-hand side to indicate that this relation holds in the limit \re{large-n}. 

The subleading terms in \re{bar-R} involve odd powers of $\hat n$ accompanied by functions $I_k(a)$ defined in \re{In}. Following \cite{Belitsky:2020qir}, we can show that the series in \re{bar-R} admits the following compact integral representation
\begin{align}\notag\label{logR}
\log \hat R_{n,\alpha} {}&= -{2\hat n\over \pi} \int_{\hat n}^\infty {dz\over z} {\log(1-s_\alpha \chi(z))\over \sqrt{z^2-\hat n^2}}+O(1/g)
\\
{}&= -{2\hat n\over \pi} \int_{\hat n}^\infty {dz\over z \sqrt{z^2-\hat n^2}}\log \bigg({ \cosh z- \cos(2\pi a)\over \cosh z-1}\bigg)+O(1/g),
\end{align}
where 
we replaced $\chi(z)$ and $s_\alpha$ by their expressions \re{eq:symbol} and \re{eq:quantities}.
The corresponding expression for the function $\hat{\mathcal V}_{n,\alpha}(g)$ can be found from \re{V} and \re{logR} as
\begin{align}\label{V-heavy}
  \hat{\mathcal{V}}_{\alpha,n} = 1-{1\over 2\pi g} \int_{\hat n}^\infty {dz\over \sqrt{z^2-\hat n^2}} {\sin^2(\pi a)\coth(z/2) \over \cosh z- \cos(2\pi a)}+O(1/g^2)\,.
\end{align}
The above relations are valid up to corrections vanishing in the limit \re{large-n}. 
Note that for $g\to\infty$ the function $\hat{\mathcal{V}}_{\alpha,n}$ ceases to depend on $\hat n$ and approaches $1$. 

The relations \re{logR} and \re{V-heavy} hold in the limit \re{large-n} for arbitrary  $a=\alpha/L$. For $a=1/2$, or equivalently $s_\alpha=1$, the relation \re{logR} coincides with the analogous relation in \cite{Beccaria:2022ypy}, see eq.~(4.41). For $a=0$ both functions coincide with their value in a free theory, $\hat{R}_{0,n} =\hat{\mathcal{V}}_{0,n} =1$, in agreement with the  properties 
 of untwisted operators.

Let us analyze the behaviour of the functions \re{logR} and \re{V-heavy} in the limit of small $a$, or equivalently $L\to\infty$. This regime is closely related to the long quiver limit \re{xi}. We therefore expect that the relations \re{logR} and \re{V-heavy} have to match \re{RV-limit} upon appropriate identification of the parameters. It is important to note that the relation \re{RV-limit} was derived at large coupling $g$ while keeping $n$ fixed. To reconcile this with the regime specified in \re{large-n}, we consider $n$ to be large but much smaller than $g$. This corresponds to the limit of small $\hat{n}$ in \re{large-n}.
 
For small $\hat n$, the dominant contribution to \re{logR} and \re{V-heavy} comes from integration around $z=\hat n$. 
We find at small $a=\alpha/L$
\begin{align}\notag\label{bar-RV}
{}&\hat R_{n,\alpha}=\left[ 2\over 1+\sqrt{1+4\xi^2/n^2}\right]^2\,,
\\
{}&\hat{\mathcal{V}}_{\alpha,n} =\left[\frac{n-1}{n}+\frac{1}{n \sqrt{1+{4 \xi ^2}/{n^2}}}\right]^{1/2},
\end{align}
where $\xi$ is defined in \re{xi}. This relation holds for $1\ll n \ll g$ and $a\ll 1$ with the ratio $\xi/n=2\pi a g/n$ held fixed. 

It is straightforward to verify that the expansion of \re{bar-RV} for small and large $\xi/n$ aligns with the relations \re{RV-small-xi} and (\ref{small-xi}), respectively, up to subleading corrections suppressed by powers of $1/n$ and $1/\xi$. For arbitrary \(\xi = O(n)\) and large \(n\), the equivalence of the relations \re{RV-limit} and \re{bar-RV} can be established by substituting the Bessel functions in \re{RV-limit} with their  expansions at large order.

\section{Correlations at strong coupling}\label{sect6} 
 
In this section, we study the long-quiver limit of the correlation function \re{OO-f} at strong coupling. 
According to \re{f-R}, the function $f_Y$ is given by the sum of $R_{\alpha,n}$ over possible values of the quasimomentum \re{quasi}.
We recall that $R_{\alpha,n}$ is protected for $\alpha=0$ and for $1\le \alpha\le L-1$ it admits the strong coupling expansion \re{RV}.
Replacing $R_{\alpha,n}$ in \re{f-R} with its leading behaviour at strong coupling \re{eq:expansion at strong}, we get
\begin{align}\label{f-naive}
f_Y={1\over L} + \frac{1}{L}\sum_{\alpha=1}^{L-1} e^{\frac{2\pi i \alpha}{L}Y} {(n-1)n\over 4g^2 s_\alpha} + O(1/g^3)\,,
\end{align}
where the first term on the right-hand side comes from $\alpha=0$.

Replacing $s_\alpha$ with its definition \re{eq:quantities}, we find that at large $L$ the sum \re{f-naive} receives the dominant contribution from $\alpha/L\ll 1$ and $(L-\alpha)/L\ll 1$. Going through the calculation, we find for $L\gg 1$ and $0\le Y\le L$ 
\begin{align}\label{f-div}
f_Y={1\over L}+ {(n-1)n L\over 4 g^2}\left[\frac{1}{3}- \frac{2 Y (L-Y)}{L^2}\right] +O(1/g^3)\,,
\end{align}
We observe that, at large $g$ and $L$, the second term on the right-hand side 
behaves as $O(L/g^2)$. Furthermore, taking into account \re{eq:expansion at strong}, we find from \re{f-R} that the subleading corrections to \re{f-div} scale as $O(L^{k}/g^{k+1})$. Consequently, the strong coupling expansion \re{f-div} is well-defined for $L/g <1$. 
In the opposite limit, for $L/g>1$, the non-perturbative corrections  in (\ref{RV})  becomes comparable to the $1/g$ corrections and the strong coupling expansion diverges.

Computing the function $f_Y$ for $L/g>1$ requires a resummation of the series (\ref{f-div}) to all orders in $1/g$. To start with, we use \re{R-inv} to rewrite the sum in \re{f-R} in a more symmetric form
\begin{align}
f_Y= \frac{1}{L}\sum_{\alpha=-L/2}^{L/2-1} e^{\frac{2\pi i \alpha}{L}Y} R_{\alpha,n}(g)\,.
\end{align}
At large $g$ and $L$, the dominant contribution to the sum comes from small $\alpha/L$.  In this region,  the function $R_{\alpha,n}$ can be replaced by its asymptotic expression  \re{RV-limit} in the limit \re{xi}, leading to  
\begin{align}\label{f-imp}
f_Y= \frac{(n-1) n}{g^2 L}\sum_{\alpha=-L/2}^{L/2-1} e^{ip_\alpha Y} \,\frac{{\mathrm I}_n(2g p_\alpha)}{p_\alpha^{2} \,{\mathrm I}_{n-2}(2g p_\alpha)}\,.
\end{align}
The sum runs over possible values of the quasimomentum \re{quasi} in the first Brillouin zone $-\pi \le  p_\alpha< \pi$. 
Strictly speaking, the sum should be restricted to $|\alpha/L|\ll 1$, which corresponds to $|p_\alpha|\ll 1$. However, at large $g$, the contribution from $p_\alpha=O(1)$ to \re{f-imp} is suppressed by a factor of $1/g^2$ compared to contribution from small $p_\alpha$, and thus does not affect the leading behaviour of \re{f-imp}.

Note that the relation \re{f-imp} holds for large $g$ and $L$ with their ratio $L/g$ held fixed. For $L/g<1$, by isolating the $\alpha=0$ term in the sum \re{f-imp}
and
replacing the ratio of the Bessel functions for $\alpha\neq 0$ with their expansion at large argument, we arrive at \re{f-div}. For large $L/g$, the sum in \re{f-imp} becomes independent of $L$ and can be approximated by an integral over the quasimomentum
\begin{align}\label{fY-Bessel}
f_Y= \frac{(n-1) n}{g^2}\int_{-\pi}^\pi {dp\over 2\pi}  e^{ip Y} \,\frac{{\mathrm I}_n(2g p)}{p^{2} \, {\mathrm I}_{n-2}(2g p)}\,.
\end{align}
This relation is valid at strong coupling for $g\ll L$ as $L\to\infty$.  

\begin{figure}[t!]
	\begin{center}
				\includegraphics[width=0.65\textwidth]{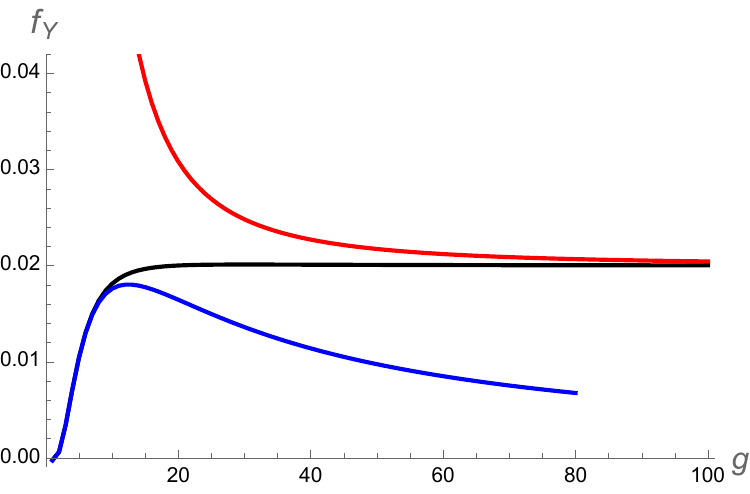}
		\caption{The dependence of the function $f_Y$ (black line) defined in \re{f-imp} on the coupling constant $1\le g\le 100$ for $L=50$, $Y=10$ and $n=2$. The red and blue lines represent its asymptotic expressions for $g\gg L$ and $g\ll L$ given by \re{f-div} and \re{fY-Bessel}, respectively. }
		\label{fy-function}
	\end{center}
\end{figure}

The dependence of the function \re{f-imp} on the coupling constant is shown in Fig.~\ref{fy-function}. We observe that it agrees with the asymptotic expressions \re{f-div} and \re{fY-Bessel} for $g \gg L$ and $g \ll L$, respectively. 

\subsection{The long quiver limit at strong coupling}

We have shown in Sect.~\ref{sect4} that, in the weak coupling regime, the function $f_Y$ decreases exponentially with the distance $Y$ in the long quiver limit $L\to\infty$. As follows from the above consideration, the properties of this function crucially depend on the ratio $L/g$. 

For $L/g \ll 1$, the function $f_Y$ is described by \re{f-div}. It depends on the quiver length $L$ and its leading $O(1/g^2)$ correction displays polynomial dependence on $Y$. We recall that in this regime the function $f_Y$ describes correlations in the lattice model in the limit when all sites are strongly correlated. 

For $L/g \gg 1$, we apply \re{fY-Bessel} and change the integration variable $p\to p/(2g)$ to find at large $g$
\begin{align}\label{fY-p-int}
f_Y=  \frac{2(n-1) n}{g}\int_{-\infty}^\infty {dp\over 2\pi}  e^{ip y} \,\frac{{\mathrm I}_n(p)}{p^{2} \,{\mathrm I}_{n-2}(p)}\,,
\end{align}
where we introduced  $y=Y/(2g)$. Note that $y$ takes finite values in the long quiver limit.

The relation \re{fY-p-int} should be compared with the analogous result \re{fY-cont} in the weak coupling regime.
In both cases, the leading large-$Y$ behaviour of the integral over $p$ is determined by the singularity of the integrand closest to the origin. 
At weak coupling, this singularity arises from a pole at $p=i \mu_n$, resulting in \re{fY-cont}. At strong coupling, the  singularities originate from the zeros of the Bessel function  ${\mathrm I}_{n-2}(p)$ within the integrand of \re{fY-p-int}. They are located along the imaginary axis at
$p=i m_{k}$, where $m_{k}$ is the $k$-th zero of the Bessel function $J_{n-2}(x)$
\begin{align}\label{m_k}
J_{n-2}(m_{k}) = 0\,,\qqqquad (k=1,2,\dots )\,.
\end{align}
To avoid clutter, we do not display the dependence of $m_k$ on $n$.

The ratio of the Bessel functions in \re{fY-p-int} can be expanded into an infinite sum over poles at $p=\pm i m_{k}$
\begin{align}\label{eq:ratio of Bessel Functions}
\frac{{\mathrm I}_n(p)}{p^2 {\mathrm I}_{n-2}(p)} =\sum_{k=1}^{\infty}\frac{4(n-1)}{m_{k}^2\left(p^2+m_{k}^2\right)} \, .
\end{align} 
Each term on the right-hand side contains the factor of $1/(p^2+m_{k}^2)$ which can be interpreted as an Euclidean one-dimensional propagator of a particle with momentum $p$ and mass $m_{k}$. Note that the sum in \re{eq:ratio of Bessel Functions} does not contain a massless pole $1/p^2$. 

Substituting \re{eq:ratio of Bessel Functions} into \re{fY-p-int} and closing the integration contour in the upper half-plane, we find for $y>0$
\begin{equation}
\label{eq:closed form}
f_Y = {4n (n-1)^2 \over g}
 \sum _{k=1}^{\infty } \dfrac{ {e}^{-ym_{k}} }{m_{k}^3} \,.
\end{equation}
 Here the right-hand side is given by the sum of one-dimensional propagators of particles with mass $m_k$. Being zeros of the Bessel function \re{m_k}, the masses are distinct $m_1<m_2<\dots$ and do not admit a simple closed-form expression. 
At large $y$, the sum in \re{eq:closed form} is dominated by the term involving the minimal mass $m_1$
\begin{align}\label{eq:behaviour correlator large L}
f_Y \sim e^{-y m_1} = e^{-Y m_1/(2g)}\,.
\end{align}
Remarkably, this relation exhibits the same exponential dependence on the distance between the nodes $Y$ as in the weak coupling regime \re{mu-weak}. According to \re{eq:behaviour correlator large L}, the corresponding mass scale $\mu_n$ at strong coupling is given by
\begin{align}\label{mu1}
\mu_n = {m_1\over 2g}\,.
\end{align}
Comparing this relation with \re{mu-weak}, we observe that $\mu_n$ decreases with increasing the coupling constant.

The function \re{eq:closed form} describes the correlation function \re{OO-f} of two operators placed at different nodes $Y=|I-J|$ of the quiver diagram in Fig. \ref{fig:qiuv} in the limit $L\gg g\gg 1$. In the context of the quiver lattice model, the exponential suppression of correlations \re{eq:behaviour correlator large L} at large separations $Y$  indicates that the interaction in this model is short-ranged, with $1/\mu_n=O(g)$ defining of the correlation length. This result is in an agreement with the findings of \cite{Beccaria:2023qnu}.  
   
\subsection{Limit of heavy operators}   
 
Let us examine the relation \re{eq:closed form} in the limit \re{large-n} when the scaling dimensions of the operators in \re{OO-f} become large. 

The dependence on $n$ enters the sum \re{eq:closed form} through the zeros of the Bessel function \re{m_k}. In the limit of large $n$, the zeros of the Bessel functions $J_{n-2}(x)$ have the following properties. The minimal zero $m_1$ is given by (see \cite{NIST:DLMF}, Eq.~(10.21.32)) 
\begin{align}\label{m1}
m_1=n+1.85575 \, n^{1/3} +O(n^0)\,.
\end{align}
The remaining zeros approach a nearly uniform distribution with a spacing of approximately $\pi$. Taking these properties into account we 
get from \re{eq:closed form}
\begin{align}
f_Y={4\over g}  e^{-y n }+\dots = {4\over g}  e^{-Y n/(2g)}+\dots\,,
\end{align}
where dots denote subleading corrections suppressed by powers of $1/n$ and $e^{-y}$.
The corresponding mass scale \re{mu1} approaches a finite value in the limit \re{large-n}
\begin{align}
\mu_n = {n\over 2g} = \hat n\,.
\end{align}
Thus, for the operators with large scaling dimension the correlation length $1/\mu_n$ remains finite at strong coupling. 
  
\section{Constructing a five-dimensional theory} 
\label{sec7}  

In the previous sections, we computed the function $f_Y$ determining the dependence of the correlation functions \re{OO-f} on the separation between the nodes in the quiver diagram in Fig.~\ref{fig:qiuv}. We demonstrated that in the limit of long quiver, for $L\gg g \gg 1$ with the ratio $y=Y/(2g)$ held fixed, this function approaches a finite value $f_n(y)=\lim f_Y$ given by \re{eq:closed form}. The resulting expression for the two-point function takes the form \re{eq:two-point with y}. 

The question arises whether the relation \re{eq:two-point with y} can be interpreted as a correlation function in some effective five-dimensional theory in which one additional dimension arises from the quiver diagram in the limit $L\to\infty$.  

\subsection{K\"all\'en-Lehmann representation}

As the first step in this direction, we apply \re{eq:closed form} to identify the spectrum of excitations propagating in the fifth dimension. 

Let us define a one-dimensional Euclidean propagator of a scalar particle with a mass $m$
\begin{align}
D(y,m)=\int_{-\infty}^\infty {dp\over 2\pi} {e^{ipy}\over p^2+m^2} = {e^{-|y|m}\over 2m}\,.
\end{align}
As was mentioned earlier, the function \re{eq:closed form} is given by a linear combination of these propagators. This suggests to introduce a spectral density function
\begin{align}\label{rho}
\rho_n(\mu^2)=\frac{4(n-1)}{\mu^2}\sum_{k=1}^{\infty}\delta(\mu^2-m_k^2)\,,
\end{align}
where the sum runs over zeros of the Bessel function \re{m_k}. This function satisfies the relation
\begin{align}
\int_0^\infty d\mu^2 \rho_n(\mu^2)=1\,,
\end{align}
which can be obtained from the identity \re{eq:ratio of Bessel Functions} by examining the leading asymptotic behaviour on both sides for $p\to\infty$.

Combining the above relations, we obtain the K\"all\'en-Lehmann representation of the function \re{eq:closed form}
\begin{align}
f_n(y)={2n (n-1)\over g}\int_0^\infty d\mu^2 \rho_n(\mu^2) D(y,\mu)\,.
\end{align}
Substitution of this relation into \re{eq:two-point with y} results in a five-dimensional representation of the two-point function $\langle O_n(x_1,y_1) \bar{O}_n(x_2,y_2) \rangle$ in the long quiver limit. 

The observed factorization of the correlation function \re{eq:two-point with y} into a product of functions solely dependent on the four-dimensional coordinates, $x$, and the fifth-dimensional coordinate, $y$, implies the decoupling of the corresponding excitations. Consequently, the excitations propagating in the fifth dimension acquire mass, while the four-dimensional excitations remain massless, thereby preserving the conformal invariance of the four-dimensional theory.

\subsection{Three-point functions}

So far our discussion was restricted to the two-point functions \re{OO-f}. Let us extend analysis to the three-point function \re{OOObar ft}.

Applying a discrete Fourier transform \re{T-def} to each operator in \re{G3}, we obtain the following representation for the three-point function 
\begin{align}\label{OOOb} 
\vev{O_{n_1}(x_1,y_1) O_{n_2}(x_2,y_2) \bar O_{n_3}(x_3,y_3)} = \cG_{n_1n_2n_3} \dfrac{\sqrt{n_1 n_2 n_3}}{N} {f_{\bm n}(y_1,y_2,y_3)\over |x_1-x_3|^{2n_1} |x_2-x_3|^{2n_2}}\,.
\end{align}
The normalization factor $\cG_{n_1n_2n_3}=\delta_{n_1+n_2,n_3}\sqrt{\cG_{n_1}\cG_{n_2}\cG_{n_3}}$ depends on the analogous factors defined in \re{eq:coefficients} and it  is nonzero only when $n_3=n_1+n_2$. The function $f(y_1,y_2,y_3)$ depends on the separations between the operators $y_{ij}=y_i-y_j$ and their scaling dimensions $\bm n=(n_1,n_2,n_3)$. In a free theory, for zero coupling constant, it equals $1$ for $y_1=y_2=y_3$ and vanishes otherwise.  For nonzero coupling constant, this function is given by (see \re{G123-F})
\begin{align}\label{f123}
f_{\bm n}(y_1,y_2,y_3)={1\over L^{2}}\sum_{\alpha_1,\alpha_2=0}^{L-1} e^{{2\pi i\over L}(\alpha_1 Y_{13} +\alpha_2 Y_{23} )} \prod_{i=1}^3
e^{\mathcal F_{n_i}-\mathcal F_{n_i-1}}\,,
\end{align}
where $Y_{ij}=I_i-I_j=2g(y_i-y_j)$. The functions $\mathcal F_{n_i}$ depend on the quasimomenta $p_i=2\pi \alpha_i/L$ satisfying $p_{\alpha_1}+p_{\alpha_2}+p_{\alpha_3}=0$ (mod $2\pi$). 

In the long quiver limit, the sums in \re{f123} can be replaced by integrals over the quasimomenta leading to
\begin{align}\label{f123-sym}
f_{\bm n}(y_1,y_2,y_3)= \int_{-\infty}^\infty dy_0 \, G_{n_1}(y_1-y_0) G_{n_2}(y_2-y_0) G_{n_3}(y_3-y_0)\,.
\end{align}
The function $G_n(y)$ is given by a Fourier integral
\begin{align}\label{G-def}
G_n(y) = \int_{-\pi}^\pi {dp\over 2\pi}e^{ip Y} e^{\widetilde{\mathcal F}_{n}-\widetilde{\mathcal F}_{n-1}}= {n-1\over g}
\int_{-\infty}^\infty {dp\over 2\pi} \, e^{i p y} \frac{{\mathrm I}_{n-1}(p)}{p\,{\mathrm I}_{n-2}(p)}\,,
\end{align}
where in the second relation we replaced $\widetilde{\mathcal F}_{n}$ with its expression \re{F-limit}, changed the integration variable as 
$p\to p/(2g)$ and took the limit of large $g$.
The integral in \re{f123-sym} is symmetric under the exchange of any pair of $(y_i,n_i)$ and $(y_j,n_j)$.  
It is interesting to note that the two-point function \re{fY-p-int} admits a representation similar to \re{f123-sym}  
\begin{align}\label{f2}
f_n(y_1-y_2)=2g \int_{-\infty}^\infty dy_0\, G_n(y_1-y_0) G_{n+1}(y_2-y_0)\,.
\end{align}
It follows from the analogous relation between the integrands of \re{fY-p-int} and \re{G-def}.

The integral in \re{G-def} is similar to that in \re{fY-p-int} and it can be evaluated in the same manner. We use the properties of the Bessel function to replace in \re{G-def} 
\begin{align}
{{\mathrm I}_{n-1}(p)\over p {\mathrm I}_{n-2}(p)}=\sum_{k\ge 1}  {2\over p^2+m_{k}^2}\,,
\end{align}
where the sum runs over zeros of the Bessel function \re{m_k}. In this way, we obtain the following representation of the function \re{G-def}
\begin{align}\notag
G_n(y) {}&= {2(n-1)\over g} \sum_{k\ge 1} \int_{-\infty}^\infty {dp\over 2\pi}{e^{ipy}\over p^2+m_k^2}
\\
{}& ={n-1\over g}\sum_{k\ge 1}{e^{-m_{k} |y|}\over m_{k}} = {1\over 2g}\int_0^\infty d\mu\, \mu^2\,  \rho_n(\mu^2) D(y,\mu)\,,
\end{align}
where the spectral density is given by \re{rho}. 

Similar to \re{eq:behaviour correlator large L}, the leading behaviour of $G_n(y)$ at large $y$ is determined by the minimal zero $m_1$ of the Bessel function \re{m_k}
\begin{align}\label{G-as}
G_n(y) \sim {n-1\over g \nu_n} e^{-\nu_n  |y|} \,,
\end{align}
where $\nu_n =m_1$  depends on $n$ (see \re{m1}). Substituting this relation into \re{f2} we find that the two-point function decays exponentially fast at the last distances $f_n(y)\sim e^{-\nu_n |y|}$. In the similar manner, the three-point function \re{f123-sym}
is rapidly decreasing function for $y_1\gg y_2,y_3$
\begin{align}
	\label{eq:fy123 cont}
f_{\bm n}(y_1,y_2,y_3)\sim e^{-\nu \, y_1} \,,
\end{align}
where $\nu={\rm min}(\nu_{n_1}, \nu_{n_2}+\nu_{n_3})$. 

The relations \re{G-as} and \re{eq:fy123 cont} are in agreement with our expectations that, due to a finite mass of excitations propagating in the fifth dimensions, the correlation functions should be exponentially decreasing functions of the separations of the operators.  


   
\subsection{Dual holographic description}   

Let us compare our results for the strong coupling regime of the ${\sf Q}_L$ theory with those obtained previously using  the AdS/CFT correspondence. 

The correlation functions \re{eq:2p intro} are holographically described by the dynamics of fields propagating in the AdS space \cite{Aharony:1999ti}. The effective action for the supergravity modes dual to the operators $T_{\alpha,n}(x)$ was derived in \cite{Gukov:1998kk,Billo:2022gmq,Billo:2022fnb,Skrzypek:2023fkr}. Using this effective action, the two- and three-point correlation functions \re{eq:2p intro} were computed in \cite{Billo:2022fnb} in terms of effective Witten diagrams. The resulting expression for the OPE coefficients \re{eq:strcutre constant def} is \footnote{Strictly speaking, this relation only holds for the twisted operators with $\alpha_i\neq 0$. For the untwisted operators with $\alpha_i=0$ one has to replace $\sqrt{n_i-1}\to \sqrt{n_i}$ in \re{C123-hol}. }
\begin{align}\label{C123-hol}
 C_3= \dfrac{\sqrt{(n_1-1)(n_2-1)(n_3-1)}}{\sqrt{L}N} +O(1/g)\,,
\end{align}
where the last term describes the subleading correction coming from the stringy modes. This relation aligns with the analogous result \re{eq:strcutre constant def}, derived from localization, after replacing the functions $\mathcal{V}_{\alpha_i,n_i}(g)$ with their leading behavior at strong coupling \re{RV}.

An important question is whether \eqref{C123-hol} provides a good approximation for $C_3$ at strong coupling. We observe that the leading term in \re{C123-hol} is independent of the quasimomenta $p_{\alpha_i}=2\pi \alpha_i/L$ of the operators. This dependence arises as soon as we take into account the subleading corrections to \re{C123-hol} in $1/g$. We have demonstrated in the previous sections that for small values of the quasimomenta, or equivalently in the large $L$ limit, these corrections run in powers of $1/(gp_\alpha)$. As a result, for sufficiently long quiver, the subleading corrections become comparable with the leading term in \re{C123-hol}, thus invalidating the strong coupling expansion of $C_3$. 

Applying \re{RV-limit} we find that the OPE coefficients \re{eq:strcutre constant def} are given in the long quiver limit by 
\begin{align}\label{C123-resum}
 C_3= \dfrac{\sqrt{(n_1-1)(n_2-1)(n_3-1)}}{\sqrt{L}N}\prod_{i=1}^3 \left[\frac{{\mathrm I}^2_{n-1}(2\xi_i)}{{\mathrm I}_{n-2}(2\xi_i) {\mathrm I}_n(2\xi_i)}\right]^{1/2},
\end{align}
where $\xi_i=2\pi \alpha_i g/L$. As explained earlier, the last factor in this relation takes into account both perturbative corrections in $1/g$
as well as nonpertubative exponentially small corrections. 

The relation \re{C123-hol} corresponds to the limit of \re{C123-resum} when all $\xi_i$ are send to infinity. Another interesting feature of the relation \re{C123-resum} is that for $\alpha_i\to 0$ (or equivalently $\xi_i\to 0$)  it correctly reproduces the OPE coefficient for the untwisted operators
\begin{align}\label{C3-free}
\lim_{\alpha_i\to 0}  C_3= \frac{\sqrt{n_1n_2n_3}}{\sqrt{L}N}\,.
\end{align}
In the dual holographic description, the last factor in \re{C123-resum} should arise from the effective action of the Kaluza-Klein (KK) mode $\eta_{\alpha_i,n_i}$ dual to the operators $T_{\alpha_i,n_i}(x)$. We can use the results of the previous sections to construct this action.

In the long quiver limit, the ${\sf Q}_L$ theory living at the boundary of the $AdS_5$ becomes effectively five-dimensional. This suggests to interpret the quasimomentum $p_\alpha$ carried by the conformal primary operator as a momentum in the fifth dimension $T_{\alpha_i,n_i}\to T_{n_i}(p_{\alpha_i})$. Consequently, the KK modes also carry this momentum $\eta_{\alpha_i,n_i}\to \eta_{n_i}(p_{\alpha_i}) $ and their effective action depends on $p_{\alpha_i}$. 

Due to the factorized dependence of the two- and three-point  correlation functions on the quasimomentum and the space-time coordinates of operators, the dynamics of the KK modes in the fifth dimension is decoupled from the one in $AdS_5$. This allows us to neglect the dependence of $T_{n}(p)$ and $\eta_{n}(p)$ on the AdS coordinates and concentrate on their dependence on $p$. 
The resulting effective action of the KK modes takes the form
\begin{align}\label{S-eff}
S_{\rm eff} = S_{\rm bdry} + S^{(2)} + S^{(3)} + \dots\,.
\end{align}
The first term describes the coupling of the local operators with the KK modes at the boundary of the AdS
\begin{align}
S_{\rm bdry}= \sum_n \int {dp\over 2\pi} \left[\bar \eta_n(p) T_{n}(-p) + \bar T_{n}(p)  \eta_n(-p) \right].
\end{align}
The two additional terms in \re{S-eff} are quadratic and cubic in the KK modes, respectively. Their general form, consistent with the conservation of $U(1)$ charge and momentum along the fifth dimension, is given by
\begin{subequations}\label{SS}
\begin{align}
{}& S^{(2)} = \sum_{n} \int   {dp \over 2\pi} \, S(p) \eta_{n}(p) \bar \eta_{n}(-p)  \,,
\\ 
{}& S^{(3)} = \sum_{n_i} \int \prod_{i=1}^3 {dp_i\over 2\pi} \, S(p_1,p_2,p_3) \eta_{n_1}(p_1)\eta_{n_2}(p_2) \bar \eta_{n_3}(p_3)\delta_{n_1+n_2,n_3}\delta(p_1+p_2+p_3) + \text{c.c.}\,,
\end{align}
\end{subequations}
where the dependence of $S(p)$ and $S(p_1,p_2,p_3)$ on the charges $n_i$ is tacitly assumed. We would like to emphasize that these relations hold for small values of the momenta $p_i$.

The $S-$functions in \re{SS} can be determined by comparing the correlators $\vev{T_{n_1}(p_1)\bar T_{n_2}(p_2)}$ and $\vev{T_{n_1}(p_1)T_{n_2}(p_2)\bar T_{n_3}(p_3)}$ computed within the effective theory \re{S-eff} with their expressions predicted by the localization in the long quiver limit. This procedure yields
\begin{subequations}\label{Ss}
\begin{align}
{}& S(p) =\omega^2 N^2 L   \frac{(n-1){\mathrm I}_n(2gp)}{(gp)^{2} \,{\mathrm I}_{n-2}(2gp)} \,,
\\ 
{}& S(p_1,p_2,p_3)=\omega^3 N^2 L  \prod_{i=1}^3  \frac{(n_i-1){\mathrm I}_{n_i-1}(2gp_i)}{g p_i\, {\mathrm I}_{n_i-2}(2gp_i)} \,,
\end{align}
\end{subequations}
where $\omega$ is a normalization factor. The OPE coefficients \re{C123-resum} are given by 
\begin{align}
C_3={S(p_1,p_2,p_3)\over \sqrt{S(p_1)S(p_2)S(p_3)}}\,.
\end{align}
We recall that the untwisted operators carry vanishing momenta along fifth dimension. The effective action of the corresponding KK modes 
$\eta_n(0)$ should be independent of the coupling constant. Indeed, we verify that  for $p_i\to 0$ both relations in \re{Ss} cease to depend on $g$ and $C_3$ coincides with its value \re{C3-free} in a free theory. For nonzero $p$ and $g\to\infty$ the function $S(0,p,-p)$ coincides (up to a normalization factor) with the analogous function defining the cubic 
term  in the effective action 
proportional to $\eta_{0,n_1}\eta_{\alpha,n_2}\bar\eta_{-\alpha,n_3}$ and 
involving one untwisted and two twisted KK modes carrying small quasimomentum $p_\alpha=2\pi\alpha/L$ (see Eq.~(5.64) in \cite{Billo:2022fnb}).

\section{Conclusions and future perspectives}\label{sect8}

In this paper, we continued the study of four-dimensional $\cN=2$ circular, cyclic symmetric quiver theories initiated in \cite{Beccaria:2023qnu}. In the limit of long quiver, the nodes becomes continuously distributed, suggesting a possible emergence of a five-dimensional theory through the deconstruction mechanism \cite{Arkani-Hamed:2001wsh}. 
 
To investigate this, we analyzed the two- and three-point functions (\ref{eq:2p intro}) of chiral primary half-BPS operators. Using supersymmetric localization, these functions can be expressed as matrix integrals, which, in the planar limit, reduce to Fredholm determinants of semi-infinite matrices \re{eq:matrix}. This powerful representation enabled us to investigate the properties of the correlation functions across the parameter space of the quiver theory.
  
We analyzed the correlation functions in the weak and strong coupling regimes and in various limits of the number of nodes and operator scaling dimensions. We showed that their properties are naturally interpreted in terms of a one-dimensional lattice model (\ref{eq:partition function on S4}) defined on the quiver diagram (Fig.~\ref{fig:qiuv}). The emergence of a fifth dimension in the long-quiver limit, understood as the continuum limit of this lattice model, crucially depends on the hierarchy between the 't Hooft coupling $\lambda$ and the number of nodes $L$. A fifth dimension arises only in the regime $L \gg \sqrt\lambda$, where interactions in the lattice model become short-ranged. Furthermore, the conformal symmetry of the quiver theory implies that the effective five-dimensional theory is not Lorentz invariant, with the dynamics in the fifth dimension decoupled from the four-dimensional dynamics.

At weak coupling, the two- and three-point correlation functions have a well-defined scaling behaviour in the limit $L\to\infty$. They decay exponentially fast with the node separation and the characteristic mass scale grows logarithmically with the 't Hooft coupling. This scaling is consistent with the propagation of massive particles along the emergent fifth dimension.

At strong coupling, the semiclassical AdS/CFT expansion of the correlation functions diverges as $L\to\infty$. By including both perturbative corrections in $1/\sqrt\lambda$ and an infinite tower of nonperturbative, exponentially suppressed corrections, we derived a remarkably simple expression for the correlation functions in the long quiver limit. These functions exhibit exponential behaviour with node separation which is analogous to that in the weak coupling regime. We computed the spectrum of masses for particles propagating along the emergent fifth dimension and found that they are given by the zeros of Bessel functions.

It would be interesting to extend the above analysis of the long quiver limit by going beyond the planar limit. For this purpose, it is necessary to develop an efficient approach to systematically calculate the non-planar corrections to the correlators in the matrix model (\ref{eq:partition function on S4}). For the free energy and expectation value of circular Wilson loop this was done in \cite{Beccaria:2021ksw,Beccaria:2023kbl,Beccaria:2023qnu}. The analytical control of the nonplanar corrections can shed new light on the holographic description of the emergent five-dimensional theory describing long quivers.

Another interesting direction is the study of high-point correlation functions of the conformal primary operators \re{T-def}. Unlike two- and three-point functions \re{eq:2p intro}, their dependence on four-dimensional coordinates is not fixed by conformal symmetry. In the long-quiver limit, these correlation functions depend on both four-dimensional cross-ratios and five-dimensional momenta. Determining these functions at strong coupling and interpreting them within the emergent five-dimensional theory remain challenging.
 
 \section*{Acknowledgments}
 
We thank M. Billo', L. Griguolo, I. Kostov, A. Lerda, D. Serban,  D. Maz\'a\v{c} and K. Zarembo for helpful discussions. We are also grateful to A. Tseytlin for a careful reading of the manuscript and very useful comments. The work of GK was supported by the French National Agency for Research grant ``Observables'' (ANR-24-CE31-7996). A.T. is grateful to the Institut de Physique Théorique for the kind hospitality during part of this work.

\bibliographystyle{JHEP}
\bibliography{Referenze}

\end{document}